\documentclass[12pt]{article}

\usepackage{amsmath}         
\usepackage{amssymb}         
\usepackage{amsfonts}        
\usepackage{graphicx}        

\usepackage{color}         
\usepackage{slashed}       
\usepackage{framed}        
\usepackage{subcaption}    
\usepackage{mathrsfs}      
\usepackage{ragged2e}      
\usepackage{lipsum}        
\usepackage{paralist}      
\usepackage{multirow}      
\usepackage{cite}          
\usepackage{booktabs}      
\usepackage{nicefrac}      
\usepackage{youngtab}	   
\usepackage{arydshln} 	   
\usepackage{appendix}      
\usepackage{setspace}      

\usepackage[font={small, it}]{caption} 

\usepackage{pifont}        
\usepackage{bbm}           
\usepackage{dtklogos}      
\usepackage{tocloft}       

\usepackage{subcaption}     

\usepackage{bibspacing}
\usepackage[T1]{fontenc}
\usepackage{longtable}
\usepackage{xhfill}

\usepackage{flip-acronyms} 

\usepackage[margin=2cm]{geometry} 

\graphicspath{{figures/}}	    
\numberwithin{equation}{section}

\renewcommand{\tilde}{\widetilde}

\newenvironment{institutions}[1][2em]
  {\begin{list}{}{\setlength\leftmargin{#1}\setlength\rightmargin{#1}}\item[]\RaggedRight}
  {\end{list}}
  
\makeatletter
\renewcommand{\@cftmaketoctitle}{} 
\makeatother
\makeatletter
\renewcommand{\cftsecpresnum}{\begin{lrbox}{\@tempboxa}}
\renewcommand{\cftsecaftersnum}{\end{lrbox}}
\makeatother

\usepackage[
	colorlinks=true,
	citecolor=black,
	linkcolor=black,
	urlcolor=blue,
	hypertexnames=false]{hyperref}

\begin{document}

\thispagestyle{empty}
\begin{center}

	{	
		\huge \bf 
		Hidden On-Shell Mediators for 
		\\
        the Galactic Center $\gamma$-ray Excess
		\\
		\vspace{.1em}
	}

\vskip .7cm



\renewcommand{\thefootnote}{\alph{footnote}} 

{	\bf
	Mohammad Abdullah\footnote{\tt
	\href{mailto:maabdull@uci.edu}{maabdull@},
		 \;
	$^b$\href{mailto:adifranz@uci.edu}{adifranz@},
		 \;
	$^c$\href{mailto:arajaram@uci.edu}{arajaram@},
		 \;
	$^d$\href{mailto:ttait@uci.edu}{ttait@},
		 \;
	$^e$\href{mailto:flip.tanedo@uci.edu}{flip.tanedo@},
		 \;
	$^f$\href{mailto:awijangc@uci.edu}{awijangc@}
	uci.edu
 }, \,
	Anthony DiFranzo$^{b}$, \,
	Arvind Rajaraman$^{c}$,\\
	Tim M.\ P.\ Tait$^{d}$, \,
	Philip Tanedo$^{e}$, \,
	Alexander M.\ Wijangco$^{f}$,
}  

\renewcommand{\thefootnote}{\arabic{footnote}}
\setcounter{footnote}{0}

\vspace{-.3cm}

\begin{institutions}[2.25cm]
\footnotesize

{\it Department of Physics \& Astronomy, University of California, Irvine, \textsc{ca} 92697} 

\end{institutions}

\end{center}


\begin{abstract}
\noindent 
We present simplified models for the galactic center $\gamma$-ray excess where Dirac dark matter annihilates into pairs or triplets of on-shell bosonic mediators to the Standard Model.
These annihilation modes allow the dark matter mass to be heavier than those of conventional effective theories for the $\gamma$-ray excess.
Because the annihilation rate is set by the dark matter--mediator coupling, the Standard Model coupling can be made parametrically small to `hide' the dark sector by suppressing direct detection and collider signals.
We explore the viability of these models as a thermal relic and on the role of the mediators for controlling the $\gamma$-ray spectral shape. We comment on ultraviolet completions for these simplified models and novel options for Standard Model final states.
\end{abstract}


\vspace{1em}



\setcounter{tocdepth}{1}
\tableofcontents
\vspace{1em}


\section{Introduction}

\begin{figure}[t]
    \centering
    \begin{subfigure}[b]{0.25\textwidth}
        \includegraphics[width=\textwidth]{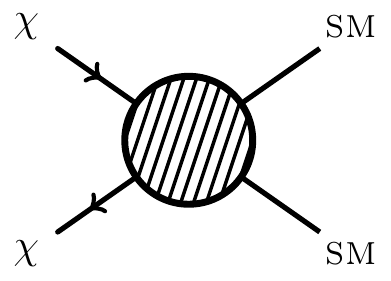}
        \caption{}
    \end{subfigure}
    \qquad\quad
    \begin{subfigure}[b]{0.25\textwidth}
        \includegraphics[width=\textwidth]{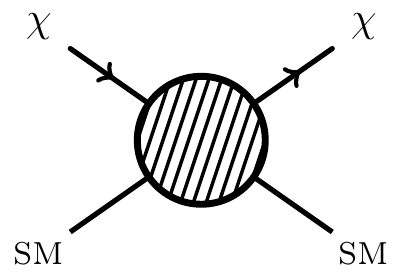}
        \caption{}
    \end{subfigure}
    \qquad\quad
    \begin{subfigure}[b]{0.25\textwidth}
        \includegraphics[width=\textwidth]{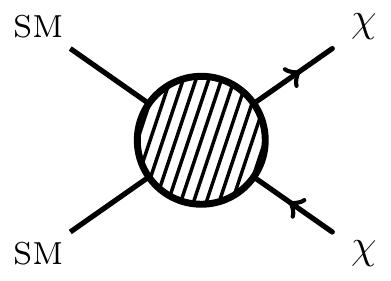}
        \caption{}
    \end{subfigure}
    
\caption{(a) Annihilation, (b) Direct Detection, (c) Collider. Complimentary modes of dark matter detection. Annihilation sets both the thermal relic abundance and the present-day indirect detection rate.}
\label{fig:blob:diagrams}
\end{figure}

The particle nature of dark matter (\DM) remains one of the outstanding open questions in high energy physics. Experimental probes of the dynamics that connect the dark sector and the Standard Model (\SM) fall into three complimentary classes shown schematically in Fig.~\ref{fig:blob:diagrams} See \cite{Arrenberg:2013rzp} for a status report. 

Recent analyses of the \FERMI Space Telescope data find an excess of 1--10 \GeV $\gamma$-rays from the center of the galaxy.
In fact, a similar excess seems to extend away from the center to high galactic latitudes~\cite{Hooper:2013rwa, Okada:2013bna, Huang:2013pda}.
This may be indicative of dark matter annihilating into \SM final states which later shower to produce the observed excess photon spectrum \cite{Goodenough:2009gk, Hooper:2010mq, Abazajian:2010zy, Abazajian:2012pn, Boyarsky:2010dr, Hooper:2011ti, Gordon:2013vta, Macias:2013vya, Abazajian:2014fta, Daylan:2014rsa, Modak:2013jya}; see \cite{Anchordoqui:2013pta,Ko:2014gha,Buckley:2011mm,Boucenna:2011hy,Zhu:2011dz,Kyae:2013qna, Marshall:2011mm,Cerdeno:2014cda,Modak:2013jya, Agrawal:2014una} for recent models. 
While an early estimate argued that an alternate interpretation based on unidentified millisecond pulsars is unlikely \cite{Hooper:2013nhl}, \cite{Abazajian:2014fta} and \cite{Yuan:2014rca} recently demonstrated the consistency of this hypothesis with the $\gamma$-ray excess. Indeed,  it may be difficult to distinguish these two possibilities since the extrapolated millisecond pulsar (\textsc{msp}) profile is very similar to standard \DM profiles  \cite{UCI_MSP}.
For the remainder of this paper we assume the excess is generated by \DM annihilation.
The latest analyses prefer a 40~\GeV dark matter candidate that annihilates into $b\bar b$ pairs\footnote{%
Annihilation of 10~\GeV \DM into $\tau\bar\tau$ is also plausible fit, see \cite{Hagiwara:2013qya,Buckley:2013sca,Buckley:2011mm,Boucenna:2011hy,Marshall:2011mm,Logan:2010nw} for recent models.  
\cite{Lacroix:2014eea} found that a universal coupling to charged leptons may be favored after bremsstrahlung and inverse Compton scattering effects are included.
In this paper we focus on the case where the $\gamma$-ray excess is generated by $b\bar b$ pairs; we comment on more general final states in Section~\ref{sec:UV:MFV} and Appendix~\ref{app:spectral:shape}.
} 
with a thermally averaged cross section $\langle \sigma v \rangle_{b\bar b} \approx \mathcal O(\text{few}) \times 10^{-26} \text{cm}^3/\text{s}$ \cite{Abazajian:2014fta, Daylan:2014rsa}.
Further, because $\langle \sigma v \rangle_{b\bar b}$ is close to the value required to be a thermal relic from standard freeze-out,
 it is implausible that such a relic could produce such a $\gamma$-ray signal without having an $s$-wave annihilation mode.
Combined with constraints from direct detection and collider experiments, this signal motivates a more detailed study of the physics encoded in the shaded regions of Fig.~\ref{fig:blob:diagrams}.

\subsection{From Effective Theories to Simplified Models}

A simple parameterization of the \SM--\DM interaction is to treat the shaded blobs as effective contact interactions between dark matter particles ($\chi$) and \SM states. For example, the coupling of fermionic \DM to a quark $q$ is parameterized through nonrenormalizable operators
\begin{align}
\mathcal L \supset \frac{1}{\Lambda^2} 
\left(\bar\chi \mathcal O_\chi \chi\right)
\, 
\left(\bar q \mathcal O_q q\right),
\label{eq:DM:EFT}
\end{align}
where, for example, $\mathcal O_\chi \otimes \mathcal O_q = \gamma^\mu \otimes \gamma_\mu$ corresponds to an interaction mediated by a heavy vector mediator that has been integrated out. The coefficient $\Lambda^{-2}$ can be calculated for specific \DM models and allow one to apply bounds from different types of experiments in a model-independent way.
This technique has been applied, for example, for collider \cite{
Cao:2009uw, 
Beltran:2010ww, 
Goodman:2010yf, 
Bai:2010hh, 
Goodman:2010ku, 
Fox:2011fx, 
Rajaraman:2011wf, 
Fox:2011pm, 
Fortin:2011hv, 
Bell:2012rg, 
Cheung:2012gi, 
Bai:2012xg, 
Ding:2012sm, 
Carpenter:2012rg, 
Cotta:2012nj, 
Zhou:2013fla, 
Carpenter:2013xra, 
Dreiner:2013vla, 
Lin:2013sca, 
Yu:2013aca, 
Berlin:2014cfa 
},
indirect detection \cite{
Beltran:2008xg,  
Goodman:2010qn, 
Cheung:2010ua, 
Cheung:2011nt, 
Rajaraman:2012fu, 
DeSimone:2013gj, 
Zheng:2010js, 
Rajaraman:2012db, 
Cheung:2012gi 
}
and direct detection \cite{
Belanger:2008sj,Kurylov:2003ra,Fitzpatrick:2012ib,Anand:2013yka,Fan:2010gt,Freytsis:2010ne,Cohen:2010gj,Gresham:2014vja,Fitzpatrick:2012ix
}
bounds on dark matter.
The choice of pairwise dark matter interactions assumes the existence of a symmetry that also stabilizes the \DM particle against decay while the pairwise \SM interactions are assumed to be the leading order gauge-invariant operators.
This need not be the case as has been demonstrated for annihilation \cite{Hochberg:2014dra} and direct detection \cite{Curtin:2013qsa}.  
In these cases, the structure in (\ref{eq:DM:EFT}) fails to capture the physics of the mediator fields which couple to both the dark and visible (\SM) sectors: the effective contact interaction description breaks down when the mediators do not decouple. The limitations of the contact interaction bounds were pointed out in \cite{Bai:2010hh} and highlighted in \cite{
Papucci:2014iwa, 
Goodman:2011jq, 
Busoni:2013lha, 
Buchmueller:2013dya 
}.

This motivates a shift in the \emph{lingua franca} used to compare experimental results to models: rather than contact interactions, light (nondecoupled) mediators suggest using `simplified models' that include the renormalizable dynamics of the mediator fields \cite{Alves:2011wf}.
This approach has been applied to
%
%
colliders \cite{
Fox:2012ru, 
Shoemaker:2011vi, 
Friedland:2011za, 
Graesser:2011vj, 
An:2012va, 
Frandsen:2012rk, 
Profumo:2013hqa, 
An:2013xka,
%
Busoni:2013lha, 
Buchmueller:2013dya, 
%
%
Goodman:2011jq, 
DiFranzo:2013vra, 
Chang:2013oia, 
Cotta:2013jna 
}
and 
astrophysical bounds where the physics of the mediator has been explored in \DM self-interactions 
\cite{Kaplinghat:2013kqa,Bellazzini:2013foa,Fan:2013yva,Tulin:2013teo,Tulin:2012wi,CyrRacine:2012fz,Foot:2012ai,Tulin:2012uq,Kaplan:2011yj,Loeb:2010gj,An:2009vq,Buckley:2009in,Feng:2009hw,Feng:2009mn,Pospelov:2008jd,ArkaniHamed:2008qn,Kesden:2006zb,Foot:2004pa,Mohapatra:2001sx, Spergel:1999mh}.

\subsection{The $\gamma$-ray Excess Suggests Light Mediators}
\label{sec:wby:light:mediators}


\begin{table}[t]
\begin{center}
    \begin{tabular}{ccl}
    \toprule 
    \textbf{Name} & \textbf{Operator} & \textbf{Constraint}\\
    \rule{0pt}{3ex}%
    D2 & $(\bar\chi\gamma_5\chi) \, (\bar q q)$ 
    & Edge of \EFT validity from monojet bounds
    \\
    D4 & $(\bar\chi\gamma_5\chi) \, (\bar q \gamma_5 q)$ 
    & Edge of \EFT validity from monojet bounds
    \\
    \rule{0pt}{3ex}%
    D5 & $(\bar\chi\gamma^\mu\chi) \,  (\bar q \gamma_\mu q)$ 
    & Spin independent direct detection    
    \\
    D6 & $(\bar\chi\gamma^\mu\gamma_5\chi) \, (\bar q\gamma_\mu q)$ 
    & Related to D5, D8 in chiral basis
    \\
    D7 & $(\bar\chi\gamma^\mu\chi) \, (\bar q \gamma_\mu\gamma_5 q)$ 
    &  Related to D5, D8 in chiral basis
    \\
    D8 & $(\bar\chi\gamma^\mu\gamma_5\chi) \, (\bar q \gamma_\mu\gamma_5 q)$ 
    & Spin dependent direct detection
    \\
    \rule{0pt}{3ex}%
    D9 & $(\bar\chi\sigma^{\mu\nu}\chi) \, (\bar q\sigma_{\mu\nu} q)$
    & Nontrivial spin-2 \UV completion
    \\
    D10 & $(\bar\chi\sigma^{\mu\nu}\gamma^5\chi) \, (\bar q\sigma_{\mu\nu} q)$
    & Nontrivial spin-2 \UV completion
    \\
    \rule{0pt}{3ex}%
    D12 & $(\bar\chi\gamma_5\chi) \, G_{\mu\nu}G^{\mu\nu}$
    & Monojet bounds
    \\
    D14 & $(\bar\chi\gamma_5\chi) \, G_{\mu\nu}\tilde G^{\mu\nu}$
    & Monojet bounds
    \\
    \bottomrule 
    \end{tabular}
\end{center}
\vspace{-1em}
\caption{Contact operators between Dirac \DM and quarks or gluons \cite{Goodman:2010ku} that support $s$-wave annihilation and the constraint for the galactic center. See \cite{Alves:2014yha} for a recent technical analysis.}
\label{tab:EFT:operators}
\end{table}

When the galactic center signal is combined with complementary bounds from direct detection and colliders, one is generically led to the limit where the contact interaction description (\ref{eq:DM:EFT}) breaks down and a simplified model description is necessary. 
By `generic' we mean no parameter tuning or additional model building is invoked. 

The tension is summarized in Table~\ref{tab:EFT:operators}, where we list the Dirac fermion dark matter contact interactions that satisfy the requirement of $s$-wave annihilation\footnote{%
Majorana dark matter relaxes these bounds by forcing some of these operators to vanish identically.
}. 
Because each effective operator simultaneously encodes the various \DM--\SM interactions in Fig.~\ref{fig:blob:diagrams}, requiring a coupling large enough to produce the $\gamma$-ray excess automatically generates signals that are constrained by null results at direct detection \cite{Akerib:2013tjd, Aprile:2013doa} and monojet \cite{ATLAS-CONF-2012-147} experiments.
These rule out operators D5, D8, D12, and D14 in Table~\ref{tab:EFT:operators}.
%
The operators D2 and D4 are at the edge of the validity of the effective theory \cite{
Busoni:2013lha, 
Buchmueller:2013dya, 
Goodman:2011jq}. 
We ignore the D9 and D10 operators since they cannot be \UV completed by a renormalizable theory.
Finally, the D6 and D7 operators are related to D5 and D8 by the chiral structure of the Standard Model. The fermionic SU(2)$_\text{L}\times$U(1)$_\text{Y}$ eigenstates are chiral so that gauge invariant interactions are naturally written in a chiral basis $\bar q \mathcal O_q P_{L,R} q$ where $P_{L,R}=\frac 12 (1\mp \gamma^5)$. Thus one generically expects that in the absence of tuning\footnote{It is worth noting that such a `coincidental' cancellation occurs in the $Z$ coupling to charged leptons which is dominantly axial due to $\sin^2\theta_W \approx 1/4$.\label{foot:lepton:Z:axial:coupling}}, the presence of vector or axial couplings implies the existence of the other.

It is thus difficult to account for the $\gamma$-ray excess in the `heavy mediator' limit where these contact interactions are valid.  
A more technical analysis of the contact interaction description was recently performed in \cite{Huang:2013apa, Cheung:2011nt, Alves:2014yha} and includes the case of scalar dark matter. 
The $\gamma$-ray excess thus generically implies a dark sector with mediators that do not decouple and hence is more accurately described in a simplified model framework.
Recent comprehensive studies of simplified models for the $\gamma$-ray excess have dark matter annihilating through off-shell mediators ($s$- and $t$-channel diagrams) \cite{Berlin:2014tja, Izaguirre:2014vva}; see \cite{Boehm:2014hva, Hektor:2014kga} for an earlier model.

\subsection{Annihilation to On-shell Mediators}

\begin{figure}
    \centering
    \begin{subfigure}[b]{0.25\textwidth}
        \includegraphics[width=\textwidth]{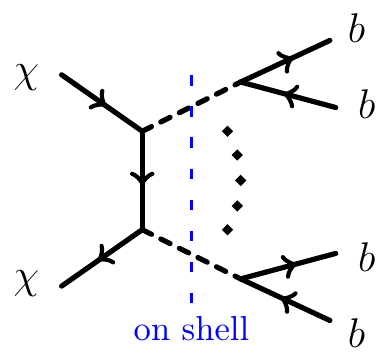}
        \caption{}
    \end{subfigure}
    \qquad\quad
    \begin{subfigure}[b]{0.25\textwidth}
        \includegraphics[width=\textwidth]{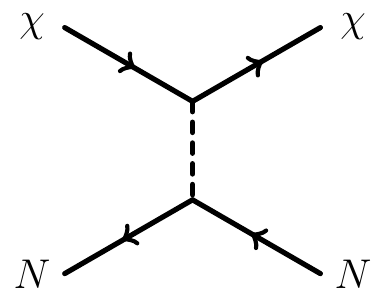}
        \caption{}
    \end{subfigure}
    \qquad\quad
    \begin{subfigure}[b]{0.25\textwidth}
        \includegraphics[width=\textwidth]{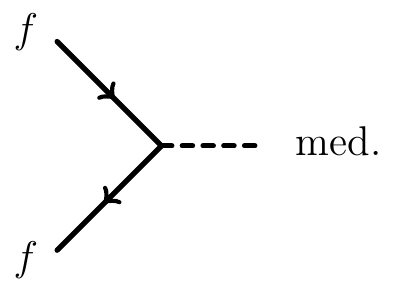}
        \caption{}
    \end{subfigure}

\caption{(a) Annihilation, (b) Direct Detection, (c) Collider. \DM complimentarity for on-shell mediators; compare to Fig.~\ref{fig:blob:diagrams}. (a) The annihilation rate is independent of the mediator coupling to the Standard Model. 
(b)  Direct detection remains 2-to-2, here $N$ is a target nucleon.
(c) Colliders can search for the presence of the mediator independently of its \DM coupling.
}
\label{fig:complimentary:mediator}
\end{figure}

In this paper we focus on a different region in the space of simplified models where mediators are light enough that they can be produced on-shell in dark matter annihilation, henceforth referred to as the on-shell mediator scenario.
This annihilation mode is largely independent of the mediator's coupling to the \SM so long the latter is nonzero.
Lower limits on the \SM coupling---that is, upper limits on the mediator lifetimes---are negligible since the mediator may propagate astrophysical distances before decaying to the $b\bar b$ pairs that subsequently yield the $\gamma$-ray excess.
The \SM coupling can be parametrically
small which suppresses the off-shell $s$-channel annihilation mode as well as the direct detection and collider signals. This is shown in Fig.~\ref{fig:complimentary:mediator}.

Because on-shell annihilation into mediators requires at least two final states\footnote{%
One may also consider semi-annihilation processes $\chi_1\chi_2 \to \chi_3 (\text{mediator})$~\cite{D'Eramo:2010ep}. See~\cite{Boehm:2014bia} for a prototype model for the galactic center $\gamma$-ray excess.
}, the resulting annihilation produces at least four $b$ quarks, as shown in Fig.~\ref{fig:complimentary:mediator}a. This, in turn, requires a heavier dark matter mass in order to eject $\approx$ 40 \GeV $b$ quarks from each annihilation to fit the $\gamma$-ray excess. This avoids the conventional wisdom that this excess requires 10 -- 40 \GeV dark matter. In the limit on-shell annihilation dominates, the total excess $\gamma$-ray flux is fit by a single parameter, the mediator coupling to dark matter. Once fit, this parameter determines whether the \DM may be a thermal relic.
We remark that the spectrum is slightly boosted by the on-shell mediator; we address this below and explore possibilities where the mediator mass can be used as a handle to change the spectral features.

The on-shell mediator limit thus separates the physics of mediators \SM and \DM couplings. The former can be made parametrically small to hide \DM from direct detection and collider experiments, while the latter can be used to independently fit indirect detection signals such as the galactic center $\gamma$-ray excess. 
Observe that these simplified models modify the standard picture of complementary \DM searches for contact interactions shown schematically in Fig.~\ref{fig:complimentary:mediator}. Annihilation now occurs through multiple mediator particles and is independent of the mediator coupling to the \SM. Direct detection proceeds as usual through single mediator exchange between \DM and \SM. Collider bounds, on the other hand, need not depend on the \DM coupling at all and can focus on detecting the mediator rather than the dark matter missing energy.

In this paper we explore the phenomenology of on-shell mediator simplified models for the galactic center.
This paper is organized as follows. In the following two sections we present the on-shell simplified models that generate the $\gamma$-ray excess and determine the range of dark sector parameters. 
We then assess in Section~\ref{sec:experimental:bounds} the extent to which the on-shell mediators must be parametrically hidden from direct detection and colliders.
In Section~\ref{sec:relic:abundance} we discuss the viability of this scenario for thermal relics.
We comment on the lessons for \UV models of dark matter in Section~\ref{sec:UV:discussion}. Appendix~\ref{app:spectral:shape} briefly describes plausible variants for generating $\gamma$-ray spectra with more diverse \SM final states.

\section{On-Shell Simplified Models}
\label{sec:simplified:models}

\begin{figure}
\begin{center}
\includegraphics[width=.75\textwidth]{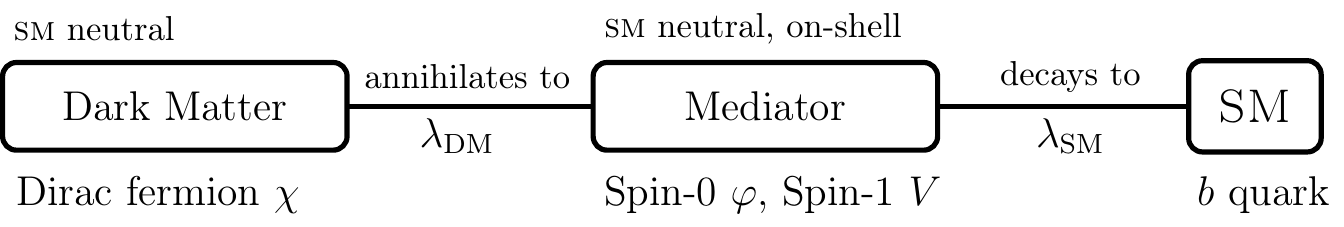}
\caption{%
Dark matter annihilates to on-shell mediators, which in turn decay into $b\bar b$ pairs. Each step is controlled be a separate coupling, $\lambda$. See text for details.
}
\label{fig:simplified:model}
\end{center}
\end{figure}

Fig.~\ref{fig:simplified:model} schematically represents the class of simplified models that we consider.
We assume the existence of a single \SM neutral spin-0 or spin-1 mediator which couples to Dirac fermion \DM with coupling $\lambda_\text\DM$ and $b\bar b$ pairs with coupling $\lambda_\text\DM$. 
Majorana fermions do not differ qualitatively in this regime.
We focus on the case where mediators couple to the Dirac \DM fermion with coupling $\lambda_\text\DM$ and to $b\bar b$ pairs with coupling $\lambda_\text\SM$. 

\subsection{Parity Versus Chirality}
\label{sec:parity:chirality}

Before describing the mediator interactions, we remark on the utility of the parity and chirality bases for four-component fermion interactions. In the parity basis, one uses explicit factors of the $\gamma^5$ matrix to parameterize
\begin{align}
\text{scalar } (\mathbbm{1}),
&&
\text{pseudoscalar } (\gamma^5),
&&
\text{vector } (\gamma^\mu),
&&
\text{and axial } (\gamma^\mu\gamma^5).
\label{eq:fermion:parity:basis}
\end{align}
interactions. This basis is most suited for nonrelativistic interactions. Equivalently, in the chirality basis, one inserts chiral projection operators
$P_{L,R} = \frac 12\left(1\mp \gamma^5\right)$
into fermion bilinears. 
This is the natural description of \SM gauge invariants.  The spin-0 fermion bilinears are 
\begin{align}
\bar \Psi (\mathbbm{1}, \gamma^5) \Psi
= 
\bar \Psi P_L \Psi \pm \bar \Psi P_R \Psi
=  \psi \chi \mp \text{h.c.}
\label{eq:fermion:bilinear:scalar}
\end{align}
where we have written the Dirac spinor in terms of two-component left-handed Weyl spinors $\Psi = (\psi,\chi^\dag)^T$, see e.g.~\cite{Dreiner:2008tw}. Similarly, the spin-1 bilinears are
\begin{align}
\bar \Psi \gamma^\mu(\mathbbm{1}, \gamma^5) \Psi
= 
\bar \Psi \gamma^\mu P_L \Psi \pm \bar \Psi \gamma^\mu P_R \Psi
=  \psi^\dag \bar\sigma^\mu \psi \mp \chi^\dag \bar\sigma^\mu \chi.
\label{eq:fermion:bilinear:vector}
\end{align}
The $\gamma^5$ appears as a phase in the spin-0 coupling and a relative sign in the spin-1 couplings of opposite chirality fermions.

The phenomenology of the $\gamma$-ray excess suggests the use of both descriptions. \DM annihilation and direct detection occur nonrelativistically so the choice of a scalar (vector) versus a pseudoscalar (axial) can dramatically affect the rate for these processes. It is thus useful to parameterize these in the language of (\ref{eq:fermion:parity:basis}), whether or not the \DM interactions are chiral. 
On the other hand, electroweak gauge invariance mandates chiral interactions for the mediator's \SM coupling.

We are thus led to consider a hybrid description where the mediator's interaction with the \SM is naturally described by a chiral coupling while the interaction with \DM is most usefully described by a coupling of definite parity. 
The chiral description of the \SM breaks down for direct detection; however, since chiral interactions generically include both the $\mathbbm{1}$ and $\gamma^5$ terms, we focus on bounds from the parity-even interaction that yields stronger bounds.
Dark matter searches at colliders probe relativistic energies without polarization information and are thus typically independent of parity. In this document we refer to the `spin-0' or `pseudoscalar' mediator to mean the spin-0 field which has a pseudoscalar interaction with the Dirac \DM without assuming a particular parity-basis interaction to the \SM.

\subsection{Mediators Versus $s$-wave Annihilation} 
\label{sec:parity:chirality:swave}

\begin{table}[t]
\begin{tabular}{rcccccc}
&

\includegraphics[width=.1\textwidth]{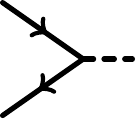}

&

\includegraphics[width=.1\textwidth]{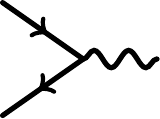}

&

\includegraphics[width=.1\textwidth]{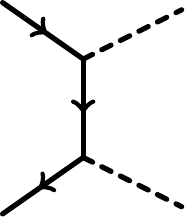}

&

\includegraphics[width=.1\textwidth]{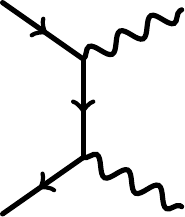}

&

\includegraphics[width=.1\textwidth]{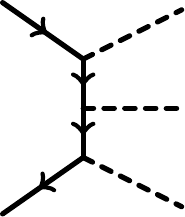}

&

\includegraphics[width=.1\textwidth]{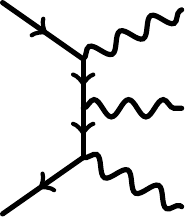}

\\
Interaction
&
\textsc s (\textsc p)
&
\textsc v (\textsc a)
&
\textsc s (\textsc p)
&
\textsc v (\textsc a)
&
\textsc s (\textsc p)
&
\textsc v (\textsc a)
\\
Partial Wave
&
$p$ ($s$)
&
$s$ ($p/s$)
&
$p$ ($p$)
&
$s$ ($s$)
&
$p$ ($s$)
&
$p$ ($p$)
\\
On/Off-Shell
&
Off
&
Off
&
On
&
On
&
On
&
On
\\
\DM Mass [\textsc{g}e\textsc{v}]
&
$\approx 40$
&
$\approx 40$
&
$\approx 80$
&
$\approx 80$
&
$\approx 120$
&
$\approx 120$
\end{tabular}
\caption{Annihilation to mediators. 
\textsc{s,p,v,a} correspond to scalar, pseudoscalar, vector, and axial vector interactions with \DM. Also shown: the leading velocity (partial wave) dependence, whether the process may occur on-shell, and the approximate mass for $40$ \GeV final state $b$ quarks.
The off-shell axial coupling is $s$- or $p$-wave for axial/vectorlike \SM coupling respectively \cite{Kumar:2013iva}.
}
\label{tab:annihilation:modes}
\end{table}

The parity basis for dark matter interactions clarifies the types of interactions that can yield $s$-wave annihilation for the $\gamma$-ray excess. 
In Table~\ref{tab:annihilation:modes} we show annihilation modes to up to three spin-0 or spin-1 mediators for the interactions in (\ref{eq:fermion:parity:basis}). On-shell kinematics require at least two final states so that the leading annihilation modes in the on-shell mediator limit are two spin-1 particles (of either parity) or three pseudoscalars.  The off-shell diagrams represent the $s$-channel simplified models in \cite{Berlin:2014tja, Izaguirre:2014vva}.

Also shown in Table~\ref{tab:annihilation:modes} are the approximate masses for the on-shell mediator scenarios. In order to eject 40 \GeV $b$ quarks from each annihilation, the two (three) body final states require that the \DM mass is approximately $m_\chi=$ 80 (120) \GeV. Observe that this mechanism allows one to circumvent the conventional wisdom that the galactic center signal requires \DM lighter than typical electroweak scale states. 

Note that these masses are back-of-the-envelope estimates that do not account for the boost in the $b$ spectrum from the mediator momentum or the spread in mediator energies for the 3-body final state. Further, we assume only couplings to $b$. This is a reasonable estimate and does not violate flavor bounds for spin-0 mediators since it follows approximately from minimal flavor violation (\MFV) \cite{Hall:1990ac,Chivukula:1987py,Buras:2000dm,D'Ambrosio:2002ex}. On the other hand, spin-1 mediators generically couple democratically to all three generations in the \MFV ansatz, as can be seen when comparing (\ref{eq:fermion:bilinear:scalar}) and (\ref{eq:fermion:bilinear:vector}). Finally, one should also account for the effect of the off-shell, $s$-channel annihilation modes for finite coupling to the \SM, $\lambda_\SM$. We account for these  in Sec.~\ref{sec:indirect:detection} where we perform a fit to the $\gamma$-ray excess.

The amplitudes for annihilation to two spin-1 mediators via the vector and axial interactions are identical so in this case the choice of parity versus chirality basis is irrelevant. 
Of the spin-0 mediators, however, only pseudoscalars generate $s$-wave annihilation. If the dark sector is described by a chiral theory, one generically expects both parities to be present. However, since the scalar is $p$-wave, it is suppressed by $\langle v^2\rangle \sim 10^{-6}$ and may be ignored for annihilation. On the other hand, this dramatically affects the direct detection rate, as discussed in Sec.~\ref{sec:sub:direct}.

%
%

\subsection{Requirements for On-Shell Mediators}

On-shell mediator models must satisfy the following conditions for the dark sector spectrum,
\begin{subequations}
\begin{align}
2 m_\chi &> \left\{
\begin{array}{ll}
2 m_V & \text{ for a spin-1 mediator}\\
3 m_\varphi & \text{ for a spin-0 mediator}
\end{array}
\right.
\label{eq:on:shell:chimass}
\\
m_{V,\varphi} & > 2m_b
\label{eq:on:shell:medmass}
\intertext{and the following requirements on the mediator couplings,}
\lambda_\text\DM & \sim 1
\label{eq:on:shell:med:req:lam:DM}
\\
\lambda_\text\SM & \ll 1.
\label{eq:on:shell:med:req:lam:SM}
\end{align}
\end{subequations}
These are interpreted as follows:
\begin{enumerate}[(a)] \itemsep-.25em
\item Nonrelativistic \DM annihilation has enough energy to produce on-shell mediators.
\item The mediator may decay into $b$ quarks to produce the spectrum of the $\gamma$-ray excess.
\item The additional coupling(s) in the on-shell diagrams do not suppress the amplitude nor are they so large that they are nonperturbative, $\lambda_\text\DM^2 < 4\pi$.
\item Parametrically suppress the off-shell, $s$-channel mediator diagrams in annihilation and simultaneously ameliorate limits from direct detection and colliders.
\end{enumerate}

We now elucidate the conditions (\ref{eq:on:shell:med:req:lam:DM}--\ref{eq:on:shell:med:req:lam:SM}) more carefully by determining the coupling scaling of the on-shell versus off-shell annihilations. 
For a spin-1 mediator, the on-shell annihilation mode goes through two on-shell mediators which subsequently decay into $b\bar b$ pairs. The key observation is that unlike the case of an off-shell $s$-channel mediator, the annihilation to on-shell mediators is largely independent of the coupling to the \SM, $\lambda_\text\SM$. We thus focus on the limit where the on-shell mode dominates over the off-shell $s$-channel diagram,
\begin{align}
\left(
\begin{aligned}
\includegraphics[width=.2\textwidth]{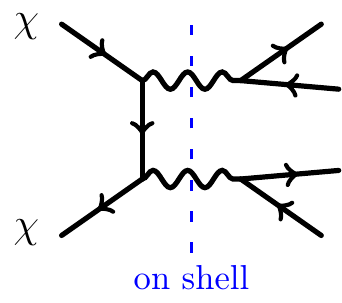}
\end{aligned}
\right)
\sim \lambda_\text\DM^2
\qquad {\Large{\gg}} \qquad
\left(
\begin{aligned}
\includegraphics[width=.2\textwidth]{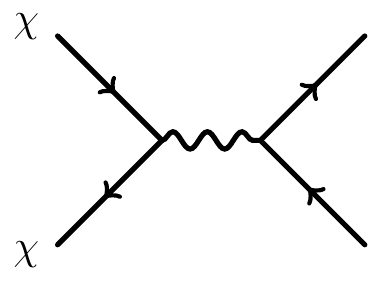}
\end{aligned}
\right)
\sim \lambda_\text{\DM} \lambda_\text{\SM}
.
\label{eq:onshell:vs:off:V}
\end{align}
Note that this condition is trivial if the mediator has axial couplings since the $s$-channel diagram is $p$-wave.
As discussed above, in a \UV model that avoids flavor bounds, a spin-1 mediator is likely to couple democratically to other \SM fermion generations. The annihilation rate relevant to the galactic center $\gamma$-ray excess would be multiplied by the branching ratio to $b\bar b$ pairs, $\text{Br}(V\to b\bar b)$. 
%
If one insists that the $\gamma$-ray excess is generated exclusively by the decay of $b$ quarks, then the branching ratio is an additional $\mathcal O(10^{-1})$ factor that must be compensated by $\lambda_\text\DM$. More dangerously, one must also account for the $\gamma$-ray pollution from annihilations yielding light quarks. We address the effect of this pollution on the fit to the $\gamma$-ray spectrum in Sec.~\ref{sec:UV:MFV}. 

For a pseudoscalar mediator the analogous limit is 
\begin{align}
\left(
\begin{aligned}
\includegraphics[width=.2\textwidth]{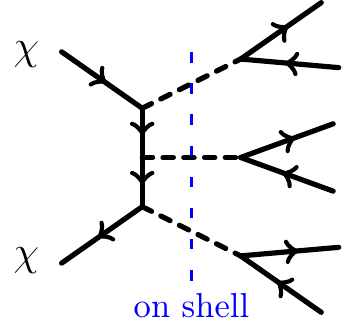}
\end{aligned}\right)
\sim \frac{\lambda_\text{\DM}^3}{\sqrt{4\pi}}
&\qquad {\Large{\gg}} \qquad
\left(
\begin{aligned}
\includegraphics[width=.2\textwidth]{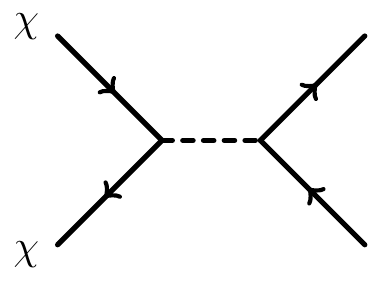}
\end{aligned}
\right)
\sim \lambda_\text{\DM} \lambda_\text{\SM}
.
\label{eq:onshell:vs:off:P}
\end{align}
We have also inserted an explicit factor of $\sqrt{4\pi}$ for the additional phase space suppression in the cross section of a three- versus two-body final state. 

Both (\ref{eq:onshell:vs:off:V}) and (\ref{eq:onshell:vs:off:P}) impose the limit $\lambda_\text{\SM} \ll 1$ to suppress the $s$-channel off-shell mediator with $\lambda_\text\DM$ fixed (for given masses) to give the correct galactic center photon yield. The magnitude of `$\gg$' is addressed in Sec.~\ref{sec:experimental:bounds}.
The limit of a very small coupling to the Standard Model is further motivated by the dearth of observational evidence for dark matter interactions at colliders and direct detection experiments. This limit also occurs naturally in models of dark photon kinetic mixing or compositeness. In our scenario, parametrically suppressing this coupling increases the lifetime of the mediator. This has little phenomenological consequence given the astronomical distance scales associated with the galactic center.

\subsection{Estimates for the $\gamma$-ray Excess}
\label{sec:estimtates:for:gce}

Before doing a fit to the $\gamma$-ray excess, we establish a back-of-the-envelope benchmark using the \DM masses in Table~\ref{tab:annihilation:modes} and neglecting the mediator spectrum and boost. This gives a reasonable estimate while also highlighting the parametric behavior of the fit.
The contact interaction fits to the galactic center $\gamma$-ray excess suggest annihilation to a pair of $b$ quarks with a thermally averaged cross section \cite{Abazajian:2014fta},
\begin{align}
\langle \sigma v \rangle_{b\bar b} \approx 5\times 10^{-26}\text { cm}^3/\text{s}.
\label{eq:Kev:bbar:xsec}
\end{align}
Note that \cite{Daylan:2014rsa} found a slightly smaller cross section, $1.5\times 10^{-26}\text { cm}^3/\text{s}$ due to a slightly tighter \DM halo (larger $\gamma$ parameter in the generalized \textsc{nfw} profile \cite{Navarro:1995iw, Navarro:1996gj, Klypin:2001xu}). The photon spectrum  from this annihilation is 
\begin{align}
\frac{d\Phi(b,\ell)}{dE_\gamma}
=
\frac{\langle \sigma v\rangle_{b\bar b}}{2} 
\frac{1}{4\pi m_\chi^2}
\frac{dN_\gamma}{dE_\gamma}
\int_\text{\textsc{los}} dx\, \rho^2\left(r_\text{gal}\left(b,\ell,x\right)\right),
\label{eq:flux:from:annihilation}
\end{align}
where $(b,\ell)$ are Galactic coordinates, $\rho$ is the \DM profile, and $r_\text{gal}$ is the distance from the galactic center along the line of sight (\textsc{los}).
%

In on-shell mediator models, the \DM annihilates 
into 2 (3) mediators which each decay into pairs of $b$ quarks. In order that each of these final state $b$ quarks to carry 40~\GeV, the \DM mass must be approximately 80 (120) \GeV as stated in Table~\ref{tab:annihilation:modes}. This reduces the \DM number density by 4 (9) in order to maintain the observed mass density; this is manifested in the $m_\chi^{-2}$ factor of (\ref{eq:flux:from:annihilation}).
This is factor is partially compensated by the multiplicity of $b\bar b$ pairs in the final state increases the total secondary photon flux by a factor of  2 (3).
Together, these effects require that the annihilation cross section is a factor of $\approx$ 2 (3) times larger than $\chi\bar\chi \to b\bar b$ cross section (\ref{eq:Kev:bbar:xsec}),
\begin{align}
\langle \sigma v \rangle_\text{ann.} \approx 2 \, (3) \times  
\langle \sigma v \rangle_{b\bar b}.
\label{eq:sigma:ann:vs:sigma:bb}
\end{align}
where $\langle \sigma v\rangle_{b\bar b}$ is the contact interaction value (\ref{eq:Kev:bbar:xsec}).
Because $\langle \sigma v\rangle_{b\bar b}$ is already determined to be close to the thermal relic, one may worry if the additional factor in (\ref{eq:sigma:ann:vs:sigma:bb}) violates the feasibility of a thermal relic. We address this in Sec.~\ref{sec:relic:abundance}. 
Considering the range of kinematically allowed mediator masses and accounting for the powers of $\lambda_\text\DM$ in the spin-0 and spin-1 cases, (\ref{eq:sigma:ann:vs:sigma:bb}) gives the estimate
\begin{align}
\lambda_\text\DM &\sim 1.1 - 1.4 & \text{ (spin-0)}
\label{eq:estimate:lambda:DM:pseudo}
\\
\lambda_\text\DM &\sim 0.27 - 0.44. & \text{ (spin-1)}
\label{eq:estimate:lambda:DM:vec}
\end{align}
These couplings indeed agree with the estimate (\ref{eq:on:shell:med:req:lam:DM}) while remaining perturbative, $\lambda_\text\DM^2 < 4\pi$. The scale of the spin-1 coupling implies a slight suppression on the left-hand side of (\ref{eq:onshell:vs:off:V}) which must be compensated by a stronger upper bound on $\lambda_\text\SM$. We show below that direct detection also constraints $\lambda_\text\SM$ strongly for the spin-1 mediator.

\begin{figure}[t]
\begin{center}
    \begin{subfigure}[b]{0.31\textwidth}
        \includegraphics[width=\textwidth]{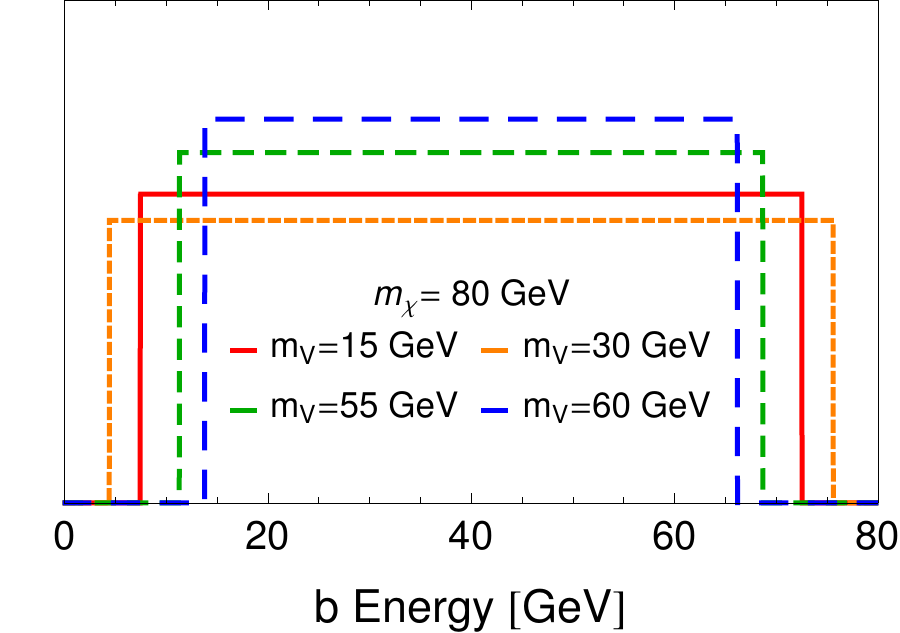}
        \caption{}
    \end{subfigure}
    \begin{subfigure}[b]{0.31\textwidth}
        \includegraphics[width=\textwidth]{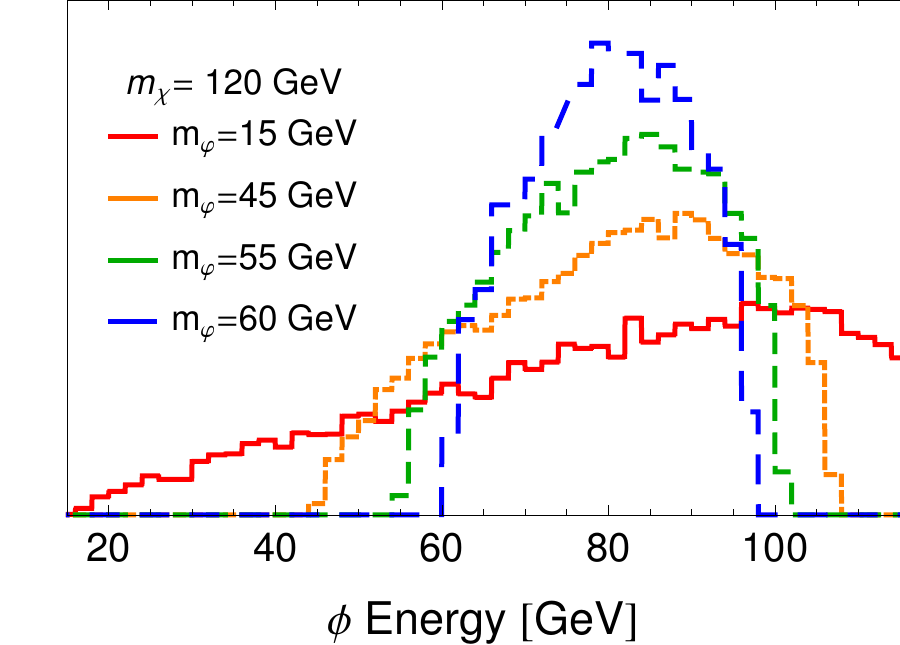}
        \caption{}
    \end{subfigure}
    \begin{subfigure}[b]{0.31\textwidth}
        \includegraphics[width=\textwidth]{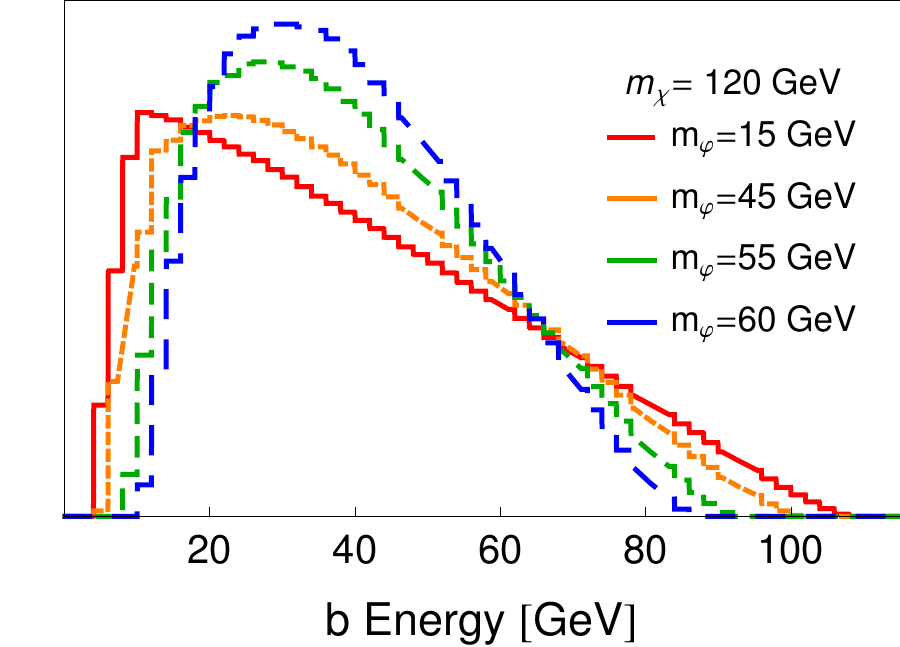}
        \caption{}
    \end{subfigure}

\caption{
(a) $\chi\bar\chi \to VV \to 4b$, (b) $\chi\bar\chi \to 3\varphi$, (c) $\chi\bar\chi \to 3\varphi \to 6b$. Energy spectrum with arbitrary normalization from \DM annihilation for (a) $b$ quarks from two on-shell spin-1 mediators, (b) pseudoscalar mediators, (c) $b$ quarks from three on-shell pseudoscalar mediators. (a) corresponds to $m_\chi = 80$ \GeV while (b,c) corresponds to $m_\chi = 120$. Lines correspond to $m_V=$ 15, 30, 55, 60 \GeV or $m_\phi =$ 15, 45, 55, 60 \GeV from red (solid) to blue (most dashed). The `box' width in (a) is not monotonically decreasing with $m_V$, as evidenced by the 30 \GeV line (orange).
}
\label{fig:mediator:spectra} 
\end{center}
\end{figure}

\section{The $\gamma$-Ray Excess from On-Shell Mediators}
\label{sec:indirect:detection}

Having established the intuition developed in Sec.~\ref{sec:estimtates:for:gce}, we examine the photon spectrum predicted from the on-shell mediator scenario and fit to the observed $\gamma$-ray excess.

\subsection{Mediator Spectra}

\begin{figure}[t]
\begin{center}
    \begin{subfigure}[b]{0.31\textwidth}
        \includegraphics[width=\textwidth]{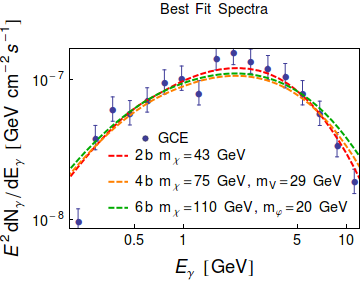}
        \caption{}
    \end{subfigure}
    \begin{subfigure}[b]{0.31\textwidth}
        \includegraphics[width=\textwidth]{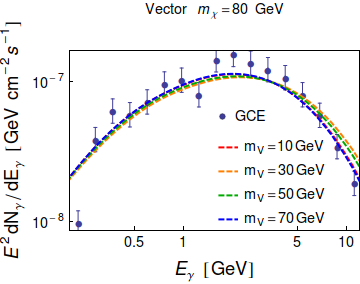}
        \caption{}
    \end{subfigure}
    \begin{subfigure}[b]{0.31\textwidth}
        \includegraphics[width=\textwidth]{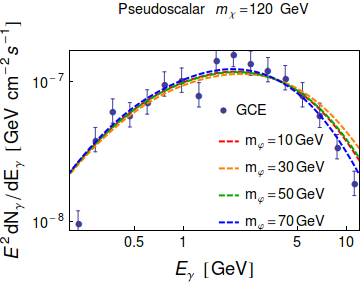}
        \caption{}
    \end{subfigure}

\caption{(a) Comparison, (b) Spin-1, (c) Spin-0. Predicted spectra for the galactic center $\gamma$-ray excess (\textsc{gce}) for (a) the best fit models categorized by the number of final state $b$ quarks, (b) a range of spin-1 mediator masses, (c) a range of spin-0 mediator masses. Overlayed is the measured $\gamma$-ray spectrum from \cite{Abazajian:2014fta}, bars demonstrate an arbitrary measure of goodness-of-fit. See Sec.~\ref{sec:fitting:spectra} for details. }
\label{fig:fits:comparison} 
\end{center}
\end{figure} 

In 2-to-2 scattering, the final state energies is completely determined by kinematics. This is the case for $\chi\bar\chi \to b\bar b$ from effective contact interactions or simplified models with single off-shell mediators; the monochromatic spectrum of final state $b$ quarks yield, upon showering, a spectrum of photons which fits the observed $\gamma$-ray excess well. 
In the case of annihilation to on-shell mediators, however, the $b$ quark spectrum is no longer monochromatic, as shown in Figure~\ref{fig:mediator:spectra}.

For spin-1 mediators, it is well known that the final states of a $\chi\bar\chi \to VV \to 4b$ cascade has box-like energy spectrum over the kinematically allowed range; see, for example, \cite{Mardon:2009rc, Ibarra:2012dw}.
The $V$ spectrum is monochromatic in the lab frame and the $b\bar b$ spectrum is monochromatic in the $V$ rest frame. The $b$ energies in the lab frame depend on the angle of the $b\bar b$ axis relative to the direction of the $V$ boost. Isotropy of the $V$ boost washes out the angular dependence and gives a flat $b$ spectrum over the kinematically allowed region.  This is demonstrated in Fig.~\ref{fig:mediator:spectra}(a). The box becomes more sharply peaked as $m_V \to m_\chi = 40$ \GeV.
The case of annihilation into three spin-0 mediators is more complicated since the mediators have a nontrivial energy spectrum and it is no longer simple to derive the $b$ spectrum from kinematics alone. Monte Carlo energy spectra for $\chi\bar\chi \to 3 \varphi$ and the subsequent decay in to $6b$ are shown in Fig.~\ref{fig:mediator:spectra}(b,c) using MadGraph~5 \cite{Alwall:2011uj}.

\subsection{Generating $\gamma$-Ray Spectra}
\label{sec:generate:gam:rays:methodology}

$\gamma$-ray spectra for our simplified models are generated using \PPPCDMID (henceforth \PPPC) \cite{PPPC4DMID, Cirelli:2010xx, Ciafaloni:2010ti}, a \textit{Mathematica} \cite{Mathematica8} package that generates indirect detection spectra based on data extracted from \PYTHIA~8 \cite{Sjostrand:2007gs}. Presently, \PPPC only generates signals for \DM annihilation into pairs of \SM particles. In order to include the effects of the on-shell mediators, one must account for the boost by convolving the \PPPC photon spectrum $dN_\gamma(E_b) /dE_\gamma$ with a distribution of $b$ energies $E_b$ which may be taken as a box for the case of two on-shell mediators or interpolated from Monte Carlo simulations such as Fig.~\ref{fig:mediator:spectra}(c).

For on-shell annihilation into spin-0 and spin-1 mediators, the shape of the photon spectrum is completely determined by the masses of the \DM particle $m_\chi$ and the mediator $m_{\varphi,V}$ while the overall normalization is fit to the necessary cross section by fixing $\lambda_\text{\DM}$, as estimated in (\ref{eq:estimate:lambda:DM:pseudo} -- \ref{eq:estimate:lambda:DM:vec}). 
The effect of the mediator mass is fairly modest, as demonstrated in the $E_\gamma^2\, dN_\gamma/dE_\gamma$ spectra in Fig.~\ref{fig:fits:comparison}. The reason for this is that the requirement that the mediator is massive enough to decay into $b\bar b$ pairs (\ref{eq:on:shell:med:req:lam:SM}) limits the extent to which the mediators are boosted.

\subsection{Fitting the $\gamma$-Ray Excess}
\label{sec:fitting:spectra}

\begin{figure}
\centering
\includegraphics[width=.4 \textwidth]{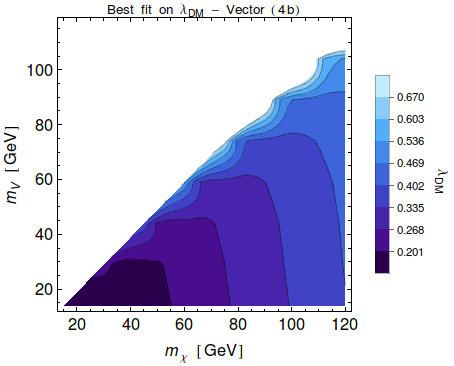}
~ 
\includegraphics[width=.4 \textwidth]{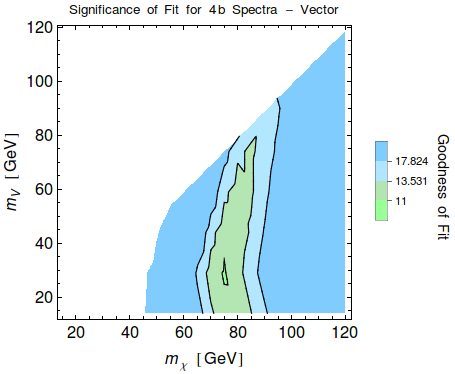}
\caption{
Fits for on-shell annihilation through spin-1 mediators. \textsc{Left}: best fit values of $\lambda_\DM$. \textsc{Right}: fit significance highlighting the best $(m_\chi, \, m_\text{med.})$ values. See text for details.}\label{fig:ID:4b} 
\end{figure}
 
We use the $\chi\bar\chi \to b\bar b$ $\gamma$-ray excess spectrum  assuming a $\chi\bar \chi \to b\bar b$ template from Figure~8 of \cite{Abazajian:2014fta}. 
We note, however, that this is an approximation since the on-shell mediator scenario predicts a different spectral shape that, in principle, should be modeled and included in the fit for the $\gamma$-ray excess. 
The comparison of the best fit $\chi\bar\chi \to 2b$ spectrum versus the on-shell mediator spectra in Fig.~\ref{fig:fits:comparison}(a) qualitatively demonstrates the degree of approximation. 

Indeed, \cite{Abazajian:2014fta} showed how the spectrum of the excess (though not its existence) can depend on both the background subtraction and the choice of \DM template assumed in the fit.
This highlights a second caveat when building \DM models for the $\gamma$-ray excess. As is standard in astrophysics literature, \cite{Abazajian:2014fta} and \cite{Daylan:2014rsa} only quote statistical errors on their fits since the systematic errors associated with fitting and subtracting background is nontrivial and intractable to quantify. Both \cite{Abazajian:2014fta} and \cite{Daylan:2014rsa} make this clear in their text. Model builders from the particle physics community, however, should be careful not to interpret these statistical uncertainties in the same way as quoted uncertainties from collider data, where both statistical and systematic errors are included. \cite{Abazajian:2014fta} demonstrated some of the systematic uncertainties by exploring the differences in the spectral fits from different background subtraction. 
Further still, both \cite{Abazajian:2014fta} and \cite{Daylan:2014rsa} use the \FERMI collaboration's 2\textsc{fgl} point sources and recommended diffuse emission model \texttt{gal\_2yearp7v6\_v0}. These assumptions also carry an implicit systematic uncertainty that are difficult to quantify without further input from the \FERMI collaboration.

That being said, one can see from the $1\sigma_\text{stat.}$ error bars in Fig.~5 of \cite{Daylan:2014rsa} that even just the statistical errors on the $\gamma$-ray excess can accommodate modified spectra. Combined with the estimated systematic errors in Figure~8 of \cite{Abazajian:2014fta} and additional systematic errors from the \FERMI background, this suggests that more general final states beyond the standard $b\bar b$ and $\tau\bar \tau$ should be considered for the $\gamma$-ray excess. In Appendix~\ref{app:spectral:shape} we present simple explorations for the range of spectra that can be generated in the on-shell mediator scenario.

Because of the unquantified systematic error associated with these spectra, we do not parameterize the statistical significance of our fits in terms of confidence intervals. Instead, we measure the goodness of fit using the $\chi^2$ value with an arbitrarily chosen 20\% error,
\begin{align}
\text{goodness of fit} = \sum_i \left( \frac{\log D_i-  \log\left(\lambda_{\text\DM}^{2n} S_i\right)}{\log(0.2 D_i)} \right)^2.
\label{eq:goodness:of:fit}
\end{align}
Smaller values are better fits.
The index $i$ runs over the bins in the extended source data set, $D$ and $S$ are the $E_\gamma^2 \frac{dN_\gamma}{dE_\gamma}$ values for the extended source data and the model spectra (assuming $\lambda_\text\DM =1$) respectively, and $\lambda_{\text\DM}^{2n}$ is the overall normalization of our input spectra, where $n=2,3$ is the number of on-shell mediators produced in each annihilation. The denominator reflects the assumed $20\%$ error: we emphasize that this is not a statement about the total error, but rather a standard candle for quantifying the goodness-of-fit. This is shown as a bar on the data in Fig.~\ref{fig:fits:comparison}.

\begin{figure}
\centering
\includegraphics[width=.4 \textwidth]{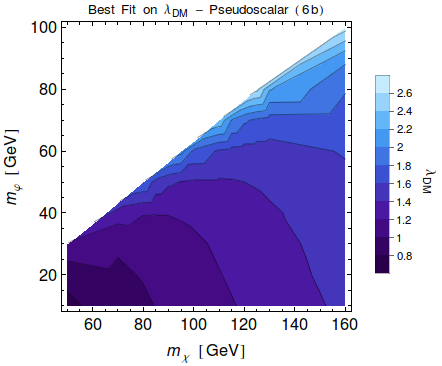}
\includegraphics[width=.4 \textwidth]{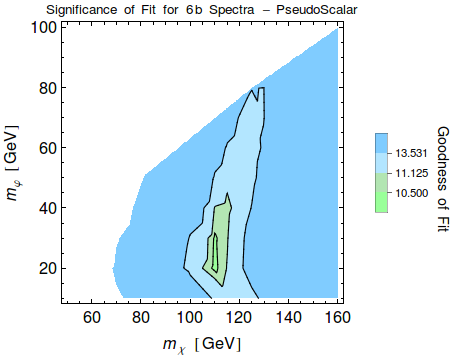}
\caption{
Fits for on-shell annihilation through spin-0 mediators. \textsc{Left}: best fit values of $\lambda_\DM$. \textsc{Right}: fit significance highlighting the best $(m_\chi, \, m_\text{med.})$ values. See text for details.}\label{fig:ID:6b} 
\end{figure}

In Figs.~\ref{fig:ID:4b} and \ref{fig:ID:6b} we fit the spectral shape over the region of \DM and mediator masses, $m_\chi$ and $m_\text{med.}$, estimated in Table~\ref{tab:annihilation:modes} and  (\ref{eq:on:shell:chimass} -- \ref{eq:on:shell:medmass}). The \DM coupling $\lambda_\text{DM}$ parameterizes the overall normalization and is fixed to minimize (\ref{eq:goodness:of:fit}) for each value of $m_\chi$ and $m_\text{med.}$.
The best fit values prefer a slightly lighter \DM particle than the back-of-the envelope estimates in in Table~\ref{tab:annihilation:modes} due to the on-shell mediator smearing the $b$ spectrum. The fits are flexible over the range of mediator masses within the kinematically accessible region, as seen in Fig.~\ref{fig:fits:comparison}(b,c).
We note that these plots assume the limit of vanishing \SM coupling, $\lambda_\text{SM}\to 0$, so that the contribution to the $\gamma$-ray spectrum from $\chi\bar\chi\to b\bar b$ via $s$-channel, off-shell mediators is negligible. We explore the role of finite $\lambda_\text{SM}$ in Sec.~\ref{sec:sub:indirect}. We also note that the simplest models spin-1 mediators typically have universal couplings to all quark generations; we address this in Sec.~\ref{sec:UV:MFV} and display the modified results in Fig.~\ref{fig:ID:4q}.


\section{Experimental Bounds on the SM Coupling}
\label{sec:experimental:bounds}

One of the features of the on-shell mediator scenario is that the $\gamma$-ray excess annihilation mode is controlled by parameters that can be independent of the conventional experimental probes for \DM--\SM interactions. Following the complimentarity in Fig.~\ref{fig:complimentary:mediator}, we examine the effect of non-negligible mediator coupling to the \SM and determine the bounds on $\lambda_\text\SM$. 

In contrast to effective contact interactions or  models with off-shell mediators, the the on-shell mediator scenario naturally includes the limit of extremely small \SM coupling so that it is always possible to parametrically `hide' from these bounds.
%
In principle, one may invoke the morphology of the $\gamma$-ray excess to set a lower bound on the mediator coupling. For example, if the mediator decay were too suppressed, the observed $\gamma$-ray excess would have a spatial extent larger than the galactic center. In fact, the \DM interpretations in \cite{Abazajian:2010zy, Daylan:2014rsa} found that the excess has a tighter profile ($\gamma>1$) than the standard \textsc{nfw} \DM density profile \cite{Navarro:1995iw, Navarro:1996gj, Klypin:2001xu}. This lower bound on $\lambda_\text\SM$ is effectively irrelevant because of the astronomical distances associated with the galactic center.

\subsection{Indirect Detection}
\label{sec:sub:indirect}


In Sec.~\ref{sec:indirect:detection} we assumed that the contribution of $s$-channel diagrams to \DM annihilation is negligible following (\ref{eq:onshell:vs:off:V} -- \ref{eq:onshell:vs:off:P}). We can use the arbitrarily normalized goodness-of-fit measure (\ref{eq:goodness:of:fit}) to assess the effect of these diagrams on the $\gamma$-ray excess fit as we parametrically increase $\lambda_\text\SM$.
We assume that the mediator couplings are such that the $s$-channel diagram supports $s$-wave annihilation, otherwise the contribution is negligible due to $p$-wave suppression by $\langle v^2\rangle \sim 10^{-6}$. From Table~\ref{tab:annihilation:modes}, we see that non-negligible $s$-channel contributions may come from mediators with either pseudoscalar or vector coupling to the \SM. 
For example, $V$ could couple axially to both \DM and the \SM with a large $s$-channel contribution for finite $\lambda_\text\SM$. On the other hand, if $V$ couples axially to \DM and vectorially to the \SM, then there may be little modification to the annihilation spectrum from $s$-channel diagrams even for large 
values of $\lambda_\text\SM$.

\begin{figure}[t]
\begin{center}
    \begin{subfigure}[b]{0.31\textwidth}
        \includegraphics[width=\textwidth]{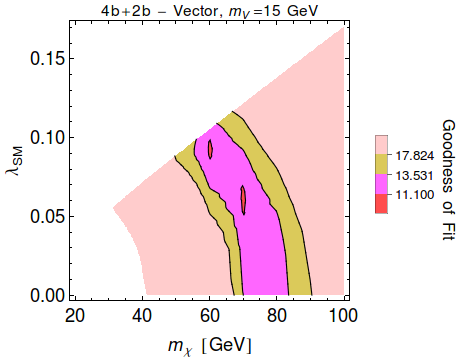}
        \caption{}
    \end{subfigure}
    \begin{subfigure}[b]{0.31\textwidth}
        \includegraphics[width=\textwidth]{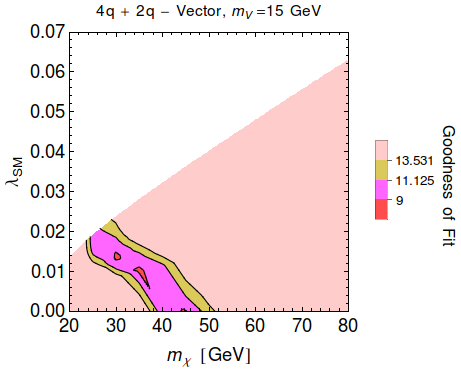}
        \caption{}
    \end{subfigure}
    \begin{subfigure}[b]{0.31\textwidth}
        \includegraphics[width=\textwidth]{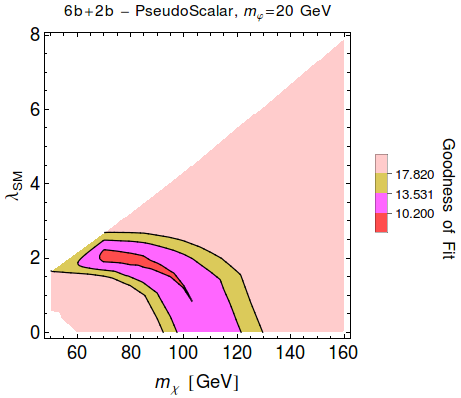}
        \caption{}
    \end{subfigure}

\caption{(a) Spin-1 ($b$-philic), (b) Spin-1 ($q$-democratic), (c) Spin-0. Fits including $s$-channel diagrams to the case of a (a) spin-1 mediator coupling only to $b$, (b) spin-1 mediator coupling to all quarks equally, and (c) pseudoscalar mediator. Plots assume that the $s$-channel diagrams are $s$-wave, see Tab.~\ref{tab:annihilation:modes}. Smaller values correspond to better fits, see (\ref{eq:goodness:of:fit}).}
\label{fig:SM:pollution} 
\end{center}
\end{figure} 

We scan over values of $\lambda_\text\SM$ that parametrically increases the relative fraction of $s$-channel off-shell \DM annihilations to on-shell annihilations to mediators\footnote{
Note from (\ref{eq:onshell:vs:off:V} -- \ref{eq:onshell:vs:off:P}) that the relative ratio of $s$-channel diagrams to on-shell mediator diagrams is determined not simply by $\lambda_\text{\SM}$, but a ratio of $\lambda_\text\SM$ to a power of $\lambda_\text\DM$ depending on the type of mediator.
}, allowing $\lambda_\text\DM$ and the mediator mass to float to a best-fit value.
The results of the fit are shown in Fig.~\ref{fig:SM:pollution}, where the best fit regions have smeared into lower \DM masses compared to Fig.~\ref{fig:ID:4b}.
The $s$-channel contribution produces $\gamma$-ray spectrum which is a poor fit due to the larger \DM mass in the on-shell mediator limit. However, because the $\gamma$-ray spectrum is smeared out relative to the $b$ spectrum, there are intermediate masses $m_\chi$ where the harder-than-usual $s$-channel diagram and the softer-than-usual on-shell mediator diagram average to yield good spectral fits. From the point of view of constructing \DM models for the $\gamma$-ray excess, this shows that not only can the \DM particle be  as heavy as 80 or 120 \GeV, as shown in Sec.~\ref{sec:simplified:models}, but it can take on intermediate values between these values and $m_\chi \approx 40$ \GeV. We further generalize this in Appendix~\ref{app:spectral:shape} where we find plausible fits with $m_\chi < 40$ \GeV, and propose a simple mechanism to make $m_\chi > 120$ \GeV.

We note that in this scenario, indirect detection bounds from cosmic antiprotons can constrain $\lambda_\text{\DM}$. Current constraints from the \textsc{Pamela} are not sensitive to the rates required in our model, though \textsc{Ams-02} will access this region \cite{Cirelli:2013hv, Kong:2014haa}\footnote{We thank \textsc{kc}~Kong for pointing this out. See Fig.~2 and Fig.~4 of \cite{Cirelli:2013hv} for the relevant bounds, recalling (\ref{eq:sigma:ann:vs:sigma:bb}) for our model. Note, however the large propagation uncertainties in Fig.~2.}.

\subsection{Direct Detection}
\label{sec:sub:direct}


\begin{figure}
\begin{center}
\includegraphics[width=.6\textwidth]{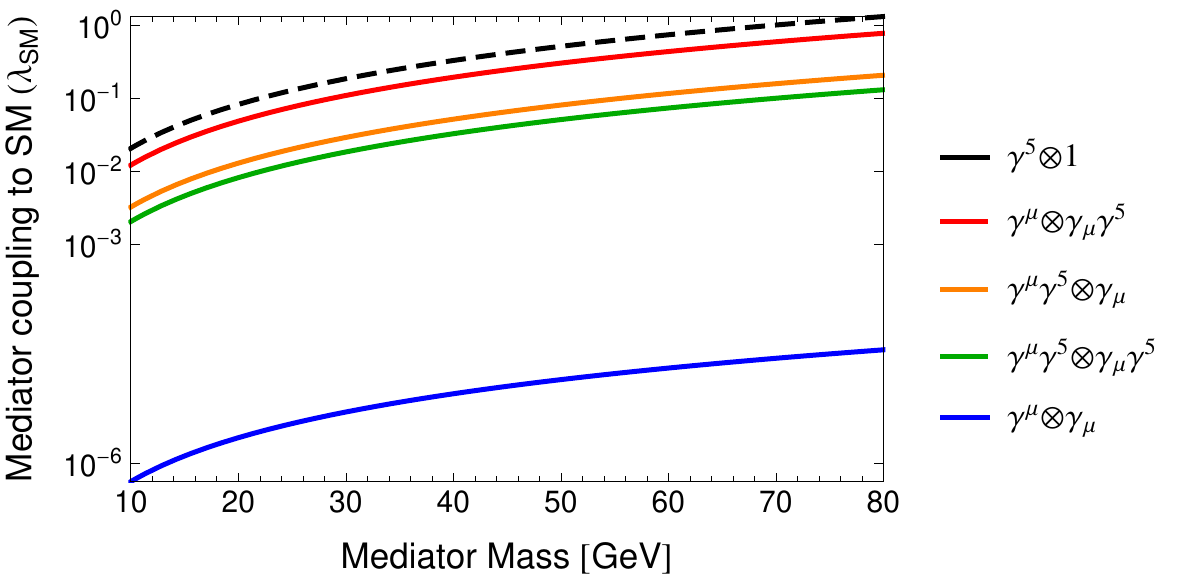}
\end{center}
\vspace{-1em}
\caption{Estimated direct detection bounds on the mediator--\SM coupling ($\lambda_\text{SM}$) for interactions $\mathcal O_\chi \otimes \mathcal O_q$ defined in (\ref{eq:DM:EFT}). The dashed (solid) lines assume the benchmark value $m_\chi =$ 120 (80) \GeV for spin-0 (1) mediators and the median \DM couplings in (\ref{eq:estimate:lambda:DM:pseudo} --\ref{eq:estimate:lambda:DM:vec}).
}\label{fig:DD}
\end{figure}

Unlike the other experimental options in Fig.~\ref{fig:complimentary:mediator},
direct detection experiments probe \WIMP--nucleon  interactions at low transfer momentum, $q^2\sim\mathcal O(10\text{ \textsc{m}e\textsc{v}})$, and are accurately described in the contact interaction limit with corrections of order $\mathcal O(q^2/m_\text{med}^2) \ll 1$. 
The present experimental bounds on the spin-independent (\textsc{si}) and spin-dependent (\textsc{sd}) interactions in the \DM mass region of interest are set by the \LUX \cite{Akerib:2013tjd} and \XENON 100 \cite{Aprile:2012nq} collaborations, respectively:
\begin{align}
\sigma_\text{\textsc{si}}
& \lesssim 10^{-45} \text{ cm}^2
&
\sigma_\text{\textsc{sd}}
& \lesssim 5\times 10^{-40} \text{ cm}^2.
\label{eq:DD:bounds}
\end{align}
In Fig.~\ref{fig:DD} we apply these bounds to the contact interactions in (\ref{eq:DM:EFT}) with the identification $\Lambda^{-2} = \lambda_\text\SM\lambda_\text\DM/m_\text{med.}^2$. We use the benchmark parameters in Section~\ref{sec:estimtates:for:gce} with the fact that the spin-0 mediator couple only to $b$ quarks while the spin-1 mediator couples universally to all quarks. 

In addition to the conventional spin-independent ($\gamma^\mu\otimes \gamma_\mu$) and spin-dependent ($\gamma^\mu\gamma^5\otimes \gamma_\mu\gamma^5$) interactions, we present bounds on the axial--vector ($\gamma^\mu\gamma^5\otimes \gamma_\mu$) and vector--axial ($\gamma^\mu\otimes \gamma_\mu\gamma^5$) interactions for a spin-1 mediator. These are suppressed by virtue of being higher order in the transfer momentum/\DM velocity; we estimate these bounds following \cite{Freytsis:2010ne}. If the spin-1 mediator couples only to $b$ quarks, the bound on $\lambda_\text\SM$ is weakened because interactions with target nucleons go through a $b$-quark loop that induces mixing between the mediator and the photon \cite{Izaguirre:2014vva, Kopp:2009et}.

As discussed in Sections~\ref{sec:parity:chirality} and \ref{sec:parity:chirality:swave}, we only consider spin-0 mediators that couple as a pseudoscalar to \DM. We do not include the $\gamma^5\otimes \gamma^5$ operator since it is so suppressed by powers of the momentum transfer that the bounds on $\lambda_\text\SM$ are weaker than the perturbativity bound $\lambda_\text{SM} < \sqrt{4\pi}$.
We evaluate momentum-dependent operators at $q^2 = 0.1 \text{ \GeV}$ following \cite{Freytsis:2010ne}.
These direct detection rates can be calculated in more detail using the nonrelativistic effective theory developed in \cite{
Fan:2010gt, 
Fitzpatrick:2012ib, 
Fitzpatrick:2012ix
}. Operator bounds in this formalism are presented in \cite{Gresham:2013mua, DelNobile:2013sia} and  \textit{Mathematica} codes for these calculations are available in \cite{DelNobile:2013sia} and \cite{Anand:2013yka}.

\subsection{Collider bounds}
\label{sec:collider}


The collider bounds for this class of models falls into two types: those based on processes where the mediator couples to both the \SM and \DM and those that only depend on the mediator's coupling to the \SM.

The first type of collider bounds are epitomized by mono-object searchers with missing energy where the \DM leaves the collider. These bounds are discussed extensively in the $\gamma$-ray [off-shell, $s$-channel] simplified models \cite{Daylan:2014rsa, Izaguirre:2014vva}. We thus only highlight the most promising proposed bound, the `mono-$b$' search \cite{Lin:2013sca}. Because of the requirement (\ref{eq:on:shell:med:req:lam:DM}) of on-shell annihilation into mediators, the class of models explored in this paper typically falls in the range where the effective contact interaction description breaks down 
\cite{Bai:2010hh, 
Papucci:2014iwa, 
Goodman:2011jq, 
Busoni:2013lha, 
Buchmueller:2013dya 
}. 
We leave a detailed simplified model study for future work, but instead translate the projected scalar--scalar ($\mathbbm{1}\otimes\mathbbm{1}$) contact interaction bounds in \cite{Lin:2013sca} as a conservative estimate for the reach of this search. Over the range of dark matter masses $m_\chi \lesssim 150$ \GeV, the projected bound from 8 \TeV \LHC data is approximately
\begin{align}
M_* > 100 \text{ \GeV}
\quad
\Rightarrow
\quad
\lambda_\text\SM^\text{spin-0} \lesssim 0.2,
\qquad
\lambda_\text\SM^\text{spin-1} \lesssim 0.6,
\label{eq:mono:b:estimated:bound}
\end{align}
where $M_*$ parameterizes the scalar--scalar contact interaction,
\begin{align}
\frac{m_q}{M_*^3} \left(\bar\chi\chi\right) \left(\bar q q\right).
\end{align}
To estimate this bound, we have matched this to $\lambda_\text{\SM}\lambda_\text{\DM} s^{-1} \left(\bar\chi\chi\right) \left(\bar q q\right)$, where we have taken $s=225$~\GeV, the cut on the minimum missing energy in \cite{Lin:2013sca}. We have estimated that the spin-1 bound on $M_*$ is identical and used the smaller $\lambda_\text\DM$ value (\ref{eq:estimate:lambda:DM:vec}). Note that at high energies the distinction between operators with and without a $\gamma^5$ in the parity basis is negligible. The bound (\ref{eq:mono:b:estimated:bound}) is thus fairly robust; unlike the direct detection bounds, a judicious choice of operator cannot avoid the constraints from this search.

A second class of collider bound comes from a search for the signatures of the mediator interacting only with the \SM sector. The bounds from this type of search are relatively weak in the mediator mass range of interest (15 -- 70 \GeV) because of large \QCD backgrounds in bump searches (dijet, $4b$); see, for example, \cite{Dobrescu:2013cmh}. Because our only requirement is that the mediator couple to $b$ quarks (and other quarks as mandated by \MFV, for example), a prototype for the mediator is a $Z'$ that gauges baryon number U(1)$_\text{B}$. This has been examined originally in \cite{Carone:1994aa, Carone:1995pu} where the most stringent bounds come from the hadronic width of the $Z$ which sets a relatively weak bound
\begin{align}
\lambda_\text{\SM} \lesssim 1.
\label{eq:Z:width:bound}
\end{align}
This bound becomes stronger in the neighborhood of the $\Upsilon$ mass, but this is already at the edge of what is kinematically allowed for decay into $b$ pairs (\ref{eq:on:shell:med:req:lam:SM}). See also \cite{Williams:2011qb} for a review including loop-level constraints from mixing and \cite{Dobrescu:2014fca} for discussion of bounds combined with anomaly constraints. 
Another prototype for the spin-0 mediator is a gauge-phobic, leptophobic Higgs. There exist very few bounds for such an object in the mass range of interest. A preliminary estimate for the reach of a `Higgs' diphoton search between 50 -- 80 \GeV  \ATLAS detector with 20/fb found weaker constraints than (\ref{eq:Z:width:bound}) \cite{Whiteson:2014gamgam}.

\section{Viability as a Thermal Relic}
\label{sec:relic:abundance}


One of the appealing features of the simplest $\chi\bar\chi \to b\bar b$ mode is that the required annihilation cross section (\ref{eq:Kev:bbar:xsec}) is so close to the value required for a thermal relic.
Due to the scaling in (\ref{eq:flux:from:annihilation}), the $s$-wave annihilation cross section for the on-shell mediator scenario is a factor of $n$ larger than the thermal value where $n=2,3$ is the number of mediators emitted, (\ref{eq:sigma:ann:vs:sigma:bb}). This comes from a factor of $n$ enhancement due to the number of $b\bar b$ final states and a factor of $n^2$ suppression coming from a decreased \DM number density. We examine the extent to which our scenario may still furnish a standard thermal relic. Observe that this sector of the model no longer has free parameters since the $\gamma$-ray excess fixes both the dark matter mass $m_\chi$ and coupling $\lambda_\text\DM$.

\subsection{$s$-wave Cross Section}


For simplicity, let us first assume that \DM annihilation at freeze-out is dominated by the same diagrams that generate the galactic center $\gamma$-ray excess at the present time. We address $s$-channel and $p$-wave corrections below. 
%
The
observed Dirac \DM density $\Omega_\chi h^2$ is approximately\footnote{The thermal cross section for Dirac \DM is a factor of 2 larger than Majorana \DM \cite{Srednicki:1988ce}.} \cite{Kolb:1994ys}
\begin{align}
\Omega_\chi h^2 \approx \frac{6 \times 10^{-27} \text{ cm}^3/s}{\langle \sigma v\rangle_\text{ann.}}
&&
\left(\Omega_\chi h^2\right)_\text{obs.} = 0.12 \quad\text{ \cite{Dunkley:2008ie, Komatsu:2008hk, Ade:2013ktc}}
\end{align}
where $h$ is the Hubble constant in units of $100 \text{ km}/(\text{s}\cdot\text{Mpc})$. 
From (\ref{eq:sigma:ann:vs:sigma:bb}), the annihilation cross section is $\langle \sigma v \rangle_\text{ann.} \approx n (5\times 10^{-26} \text{ cm}^3/\text{s})$, where $n=2$ or $3$ depending on the mediator. At face value, this gives a relic abundance that is too small. One may not mitigate this by assuming another \DM component since this, in turn, reduces the galactic center signal and hence requires one to increase the annihilation cross section further.

While the value of $\Omega_\chi h^2$ is well measured,  the precise value of the annihilation cross section $\langle \sigma v\rangle_\text{ann.}$ at freeze-out carries uncertainties from early universe parameters such as the number of effective degrees of freedom. On top of this, there are further uncertainties in our approximation (\ref{eq:sigma:ann:vs:sigma:bb}) coming from uncertainties in astrophysical parameters. For example, the $\chi\bar\chi \to b\bar b$ annihilation cross section (\ref{eq:Kev:bbar:xsec}) depends on the fit to the dark matter density profile at the center of the galaxy  \cite{Pierre:2014tra}. The analysis in \cite{Daylan:2014rsa} found a tighter density profile for which $\langle \sigma v\rangle_{b\bar b} \approx 1.5 \times 10^{-26} \text{ cm}^3/c$. 
The value of $\langle\sigma v\rangle_\text{ann.}$ spin-1 mediators ($n=2$) required for a thermal relic falls between these two estimates of  $\langle \sigma v\rangle_{b\bar b}$. We may thus assume that it is consistent with the galactic center signal within the uncertainty of the \DM morphology.
%
In fact, when the boost from the on-shell mediator is taken into account, the best fit \DM mass is slightly smaller than the assumed 80 \GeV in our estimate. This can push the estimated relic abundance from $\Omega_\chi h^2 = 0.10$ to $0.12$ 
so that the case of a spin-1 mediator may plausibly yield the correct thermal relic abundance.
On the other hand, it is difficult for a spin-0 mediator to satisfy the observed \DM relic abundance and seems to require additional mechanisms to produce $\Omega_\chi h^2$.

\subsection{$s$-channel and $p$-wave Corrections}

The corrections to the above estimates include $s$-channel $\chi\bar\chi \to b\bar b$ diagrams and $p$-wave corrections from additional on-shell mediator diagrams. The $s$-channel modes are parametrically suppressed by $\lambda^2_\text\SM \ll 1$ in the cross section and can be ignored.

Corrections from $p$-wave diagrams are negligible for present day annihilation in the galactic center due to a large velocity suppression. At the time of \DM freeze-out, on the other hand, this velocity suppression is much weaker and one should check for $p$-wave corrections to the relic abundance.
For spin-1 mediators there are no additional diagrams which are not suppressed relative to the  $\chi\bar\chi \to VV$ $s$-wave diagram.
For pseudoscalars mediators, on the other hand, the $\chi\bar\chi \to 2\varphi$ mode is $p$-wave but not parametrically suppressed by $\lambda_\text\SM$. 
At freeze-out these diagrams may contribute appreciably to \DM annihilation,
\begin{equation}
\left(
\begin{aligned}
\includegraphics[width=.2\textwidth]{fig_chichito3med}
\end{aligned}\right)
\quad\sim \quad
\frac{\lambda_\text\DM}{\sqrt{4\pi}} \sqrt{\frac{x_f}{3}}
\left(
\begin{aligned}
\includegraphics[width=.2\textwidth]{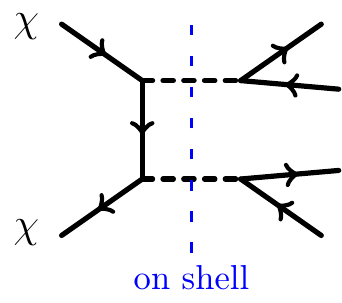}
\end{aligned}
\right)
.
\end{equation}
The prefactor accounts for the additional phase space and $p$-wave suppression. The ratio of the \DM mass to the freeze-out temperature $x_f = m_\chi/T_f \approx 20$ appears when thermally averaging the annihilation cross section at freeze-out over a Maxwell--Bolztmann velocity distribution. This factor is not especially large
and so one expects the pseudoscalar annihilation cross section at freeze-out to be even larger than approximated with only the $s$-wave piece. This further reinforces the observation that this class of mediator requires additional mechanisms to attain the observed \DM relic density. See \cite{Zurek:2013wia,Easther:2013nga,Petraki:2013wwa,McDonald:2011aa,Arcadi:2011ev,Sandick:2011rp,Hall:2009bx,Aloisio:2006yi,Feng:2004zu,Feng:2003uy,Feng:2003xh,Baltz:2001rq,McDonald:2001vt,Kolb:1998ki,Chung:1998rq,Chung:1998zb,Kofman:1994rk,Ellis:1990nb,Ellis:1984er} for a partial list of model-building tools for obtaining the correct relic abundance without the standard freeze-out mechanism.

\subsection{MSPs Can Save Freeze-Out}


As noted in the Introduction,
\cite{Abazajian:2010zy,Abazajian:2012pn,Mirabal:2013rba,Gordon:2013vta,Abazajian:2014fta, Yuan:2014rca} have pointed out that an alternate source for the $\gamma$-ray excess is a population of hitherto unobserved millisecond pulsars (\textsc{msp}s). 
As an estimate, a few thousand \textsc{msp}s could generate the observed $\gamma$-ray flux \cite{Abazajian:2014fta}. 
A recent study of low-mass X-ray binaries (\textsc{lmxb}) may lend credence to this argument. 
It is thought that \textsc{msp}s are old pulsars that have been spun up `reborn' due to mass accretion from a binary companion and that \textsc{lmxb} are simply a different phase of the same binary system. During accretion, the system is X-ray luminous and is categorized as an \textsc{lmxb}. 
The X-ray flux drops when the accretion rate drops and the system is then observed as a \textsc{msp}.
One can thus attempt to use the spatial distribution of the \textsc{lmxb} as a proxy for that of \textsc{msp}s. \cite{Voss:2007hj} found that the spatial morphology of the \textsc{lmxb} in M31 is consistent with both the $\gamma$-ray excess and the \DM interpretation---thus making it difficult to distinguish the two \cite{UCI_MSP}.

This, however, can be a boon for model-building within our \DM framework. \cite{Abazajian:2010zy} noted that the degeneracy between the \textsc{msp} and \DM intepretations of the excess suggests that the excess may come from a combination of the two sources. In this way one may  take the \DM annihilation cross section to be that which is required for a thermal relic---thus undershooting the expected $\gamma$-ray flux---and then posit that a \textsc{msp} population accounts for the remainder of the $\gamma$-ray excess.

\subsection{Conditions for Thermal Equilibrium}
\label{sec:conditions:for:th:eq}

In order for the thermal freeze-out calculation for $\chi$ to be valid, we must assume that the mediator is in thermal equilibrium when the \DM freezes out. This imposes a lower bound on the coupling of the mediator to the \SM. In principle one must solve the Boltzmann equation for the mediator, but to good approximation it is sufficient to impost $H \ll \Gamma(\text{med} \to b\bar b)$. For the range of mediators that can give the $\gamma$-ray excess, this imposes a very modest lower bound $\lambda_\text\SM \gtrsim 10^{-9}$.

\section{Comments on UV Completions and Model Building}
\label{sec:UV:discussion}

Simplified models, such as those presented here, are bridges between experimental data and explicit \UV models. In this section we highlight connections between our on-shell simplified models and viable \UV completions.

\subsection{Minimal Flavor Violation}
\label{sec:UV:MFV}

The simplified models constructed in Section~\ref{sec:simplified:models} couple the mediator only to $b$ quarks to fit to the galactic center extended $\gamma$-ray source. Assuming only this coupling violates flavor symmetry and can lead to strong constraints from flavor-changing neutral currents. A standard approach to this issue in models of new physics is to impose the minimal flavor violation (\MFV) ansatz where the Yukawa matrices are the only flavor spurions in the new physics sector \cite{Hall:1990ac,Chivukula:1987py,Buras:2000dm,D'Ambrosio:2002ex}. This prescribes a set of relative couplings to the \SM fermions up to overall prefactors. We assume that the dark sector is flavor neutral, see \cite{Batell:2011tc, Agrawal:2014una, Agrawal:2014aoa} for models with nontrivial flavor charge.

For the pseudoscalar mediator this is a small correction as can be seen by writing out the flavor indices in the spin-0 fermion bilinears (\ref{eq:fermion:bilinear:scalar}) by which the pseudoscalar couples to the quarks. MFV mandates insertions of the Yukawa matrices between couplings of right- and left-handed fermions. After rotating to mass eigenstates this yields mediator--\SM interactions 
\begin{align}
\mathcal L_\text{$\varphi$-\SM} = 
\lambda_u \frac{m_{u_i}}{\Lambda} \varphi \bar u_{Li} u_{Ri}
+ \lambda_d \frac{m_{d_i}}{\Lambda} \varphi \bar d_{Li} d_{Ri}
+ \lambda_\ell \frac{m_{\ell_i}}{\Lambda} \varphi \bar \ell_{Li} \ell_{Ri},
\label{eq:L:pseudo:MFV}
\end{align}
where $q_{L,R} = P_{L,R}q$, the $\lambda_{u,d,\ell}$ are overall prefactors, and $\Lambda$ is a \UV flavor scale. Assuming that the $\lambda_{u,d,\ell}$ are the same order naturally sets the dominant $\varphi$ decay mode to be $b\bar b$ since the $t\bar t$ mode is kinematically inaccessible for the range of masses we consider. The simplified model coupling to $b$ quarks is thus identified as
\begin{align}
\lambda_\text\SM &= \lambda_d\frac{m_b}{\Lambda}.
\end{align}
The results of the simplified model above should be adjusted by including the effects of the other $\varphi$ decay modes, though these effects are suppressed by the relative size of the other fermion masses to $m_b$.
We remark that modest to large values of $\lambda_u$ can lead to new signatures such as mediator emission off of a top quark at the \LHC or gluon couplings through top loops.

\begin{figure}
\centering
\includegraphics[width=.45 \textwidth]{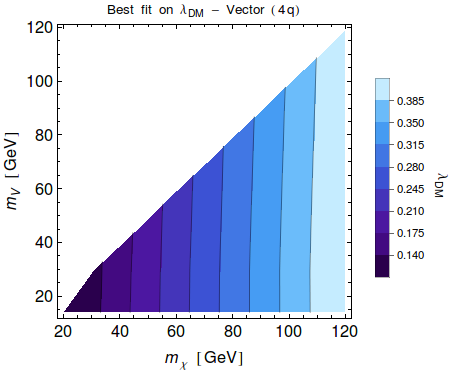}
~ 
\includegraphics[width=.45 \textwidth]{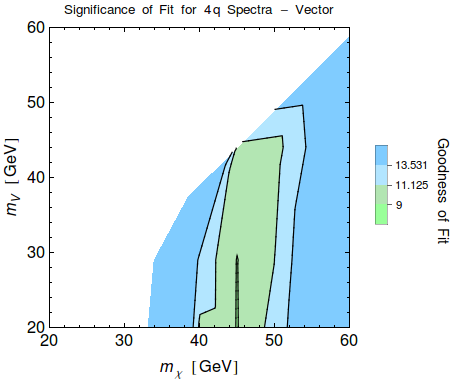}
\caption{
Fits for on-shell annihilation through spin-1 mediators assuming universal coupling to all quarks; compare to Fig.~\ref{fig:ID:4b} which assumed a coupling to only $b$ quarks.
\textsc{Left}: best fit values of $\lambda_\DM$. \textsc{Right}: fit significance highlighting the best $(m_\chi, \, m_\text{med.})$ values. See Section~\ref{sec:fitting:spectra} for details.}
\label{fig:ID:4q}
\end{figure}

The spin-1 mediators couple fermions of the same chirality, as demonstrated in (\ref{eq:fermion:bilinear:vector}). Promoting these interactions to an \MFV-compliant coupling does not introduce additional factors of the Yukawa matrices since each term is a flavor singlet. Thus, unless the \UV model is specifically constructed so that the spin-1 mediator couples preferentially to $b$ quarks, the generic expectation is the spin-1 mediators have a universal coupling to each generation, for example
\begin{align}
(\lambda_\SM)_d = (\lambda_\SM)_s = (\lambda_\SM)_b,
\end{align}
and similarly for the up-type quarks, leptons, and neutrinos. Unlike the case of the pseudoscalar mediator, this can lead to dramatic modifications since the light quarks produce a softer spectrum of secondary photons relative to the $b$. This is demonstrated in Fig.~\ref{fig:ID:4q} which shows that the best fit spectrum is very different from that of the case where the spin-1 mediator only couples to the $b$: the best fit \DM mass is $\approx 45$ \GeV rather than $\approx$ 75 \GeV.

As a caveat, we note that for fitting the $\gamma$-ray excess with either spin-0 or spin-1 mediators, it is sufficient that $\lambda_d$ is nonzero.
Thus, in principle, one can set $\lambda_u$ and $\lambda_\ell$ to vanish; the latter condition suppresses the leptonic signals for the mediator at colliders and skirts the most stringent constraints on bosons in the on-shell mediator mass range (\ref{eq:on:shell:chimass} -- \ref{eq:on:shell:medmass}).

\subsection{Gauge symmetries}

Gauge invariance also constrains \UV completions of these simplified models.
Because the \SM fermions are chiral, the parity basis spin-0 interactions on the left-hand side of (\ref{eq:fermion:bilinear:scalar}) are not SU(2)$_\text{L}\times\text{U(1)}_\text{Y}$ gauge invariant.
The similarity of (\ref{eq:L:pseudo:MFV}) to the Yukawa coupling gives a hint for how to make this interaction \SM gauge invariant.
The $m_b \bar b_R b_L$ term is implicitly $y_b (v/\sqrt{2}) \bar b_R b_L$, where $v$ is the Higgs vacuum expectation value. We may promote this to a gauge invariant coupling by restoring the Higgs doublet $H$ so that (\ref{eq:L:pseudo:MFV}) becomes
\begin{align}
\mathcal L_\text{$\varphi$-\SM}=
\frac{\lambda_u y^u_{ij}}{\Lambda}\varphi H\cdot \bar Q u_R
+
\frac{\lambda_d y^d_{ij}}{\Lambda}\varphi \tilde H\cdot \bar Q d_R
+
\frac{\lambda_\ell y^\ell_{ij}}{\Lambda}\varphi \tilde H\cdot \bar L \ell_R,
\label{eq:L:pseudo:MFV:SU2}
\end{align}
where $\tilde H = i\sigma^2 H^*$, $Q$ and $L$ are the left-handed SU(2) doublets.

\UV models for the spin-1 mediators are also constrained by gauge invariance since these couplings can be assumed to be interactions of a spontaneously broken U(1) gauge symmetry. In a \UV model one must be able to assign messenger charges to the \SM fermions---or otherwise introduce new matter in the dark sector---to cancel all gauge anomalies with respect to the mediator gauge symmetry. The axial mediator case requires particular care since the global chiral symmetry of the \SM is anomalous requiring, for example, a cancellation between the up-type and down-type quarks. See \cite{Dobrescu:2014fca} for a recent analysis of anomaly constraints on the phenomenology of $Z'$ bosons in the mass range and with the type of leptophobic/gauge-phobic couplings we consider for on-shell mediators for the $\gamma$-ray excess.

\subsection{Renormalizability}

Finally, one may push further and argue that a true `simplified model' should depend only on renormalizable couplings; i.e.\ that it should be a \UV complete theory. While the spin-1 couplings automatically satisfy this, the pseudoscalar couplings (\ref{eq:L:pseudo:MFV:SU2}) are dimension-5. We would thus like to consider renormalizable operators that generate (\ref{eq:L:pseudo:MFV:SU2}). Because the \SM fermions are chiral, there are no renormalizable interactions with the \SM singlet $\varphi$ and the \SM fermions. We thus left with interactions between the Higgs and the pseudoscalar,
\begin{align}
\mathcal L_\text{$\varphi$H}=|H|^2 \lambda_H \left(M \varphi + \varphi^2\right),
\label{eq:L:pseudo:H:portal:portal}
\end{align}
where $M$ is a dimensionful coupling. These couplings are reminiscent of the Higgs portal framework \cite{Silveira:1985rk, Patt:2006fw} with the caveat that $\varphi$ is now a mediator rather than the \DM particle. 
At energies below $m_h$, (\ref{eq:L:pseudo:H:portal:portal}) generates the couplings in (\ref{eq:L:pseudo:MFV:SU2}) with the prediction $\lambda_u = \lambda_d = \lambda_\ell$.
This is model dependent: In a two-Higgs doublet model such as the \MSSM, one may have $\varphi$ mix differently with the up- and down-type Higgses. These couplings introduce additional handles for dark sector bounds through the invisible width of the Higgs. 
See \cite{Ipek:2014gua} for an explicit model for the $\gamma$-ray excess of this type.

\subsection{Self-Interacting Dark Matter}

The on-shell mediator scenario has nontrivial dynamics even in the limit of parametrically small coupling to the \SM and may be a candidate for a model of self-interacting dark matter.
%
%
However, the lower bound on the mediator mass (\ref{eq:on:shell:med:req:lam:SM}) is heavier than the typical scale required to address anomalies in small-scale structure \cite{Tulin:2014tya,Kaplinghat:2013yxa,Kaplinghat:2013kqa,Tulin:2013teo,Tulin:2012wi,Loeb:2010gj,Buckley:2009in,Feng:2009hw,Feng:2009mn,Pospelov:2008jd,Foot:2004pa,Mohapatra:2001sx,Spergel:1999mh}. A complete study of \DM self-interactions through a pseudoscalar has yet to be completed, though the first steps are presented in \cite{Bellazzini:2013foa} and have indicated that resonance effects may be relevant even for $m_\varphi \gtrsim 10$ \GeV. Alternately, in Appendix~\ref{app:spectral:shape} we address alternate final states that may match the $\gamma$-ray excess. Of particular interest is a mediator which decays into gluons---say through a loop of heavy quarks---could be made light enough to plausibly be in the regime of interesting models for self-interaction. We leave a detailed exploration for future work.

\subsection{Prototypes for UV models}

We briefly comment on directions in specific models that may be adapted to the on-shell mediator scenario. The \MSSM introduces an additional pseudoscalar state which can plausibly mix with the Higgs as in (\ref{eq:L:pseudo:H:portal:portal}), but \SUSY bounds tend to rule out the mass range of interest. Alternately, the singlet superfield of the \NMSSM may be sufficiently unconstrained to furnish the required pseudoscalar. More generally, \cite{Ipek:2014gua} recently proposed a complete non-supersymmetric \UV model with two-Higgs doublets for the $\gamma$-ray excess. 


A second alternate direction is to develop models with spin-1 mediators. We have shown that these typically are forced to have a constrained \SM coupling if the mediator has a universal coupling to all generations, as one may generically expect for a gauged symmetry; see \cite{Hooper:2012cw} for an explicit leptophilic model. While a $Z'$ coupling to U(1)$_\text{B}$ and parametrically small coupling to the \SM is a valid scenario within the on-shell mediator framework, one may also consider options where the spin-1 mediator does not have universal coupling, for example \cite{Carone:2013uh}. 
Inspiration for such a particle is motivated by Randall-Sundrum models~\cite{Randall:1999ee} (gauge bosons with the 4D zero mode projected out, see e.g.\ \cite{Randall:2001gb, Csaki:2002gy}) or their holographic duals (composite Higgs models with $\rho$-meson-like excitations) \cite{Contino:2011np, Bellazzini:2012tv}.


\subsection{Exceptions}

Finally, we point out several exceptions to some of the `generic' statements we have made in this document.

\begin{itemize} \itemsep-.25em
\item In Sec.~\ref{sec:wby:light:mediators} we motivated the on-shell mediator scenario by exploiting how bounds on one operator `generically' bound others. Some of these bounds are avoided when $\chi$ were a Majorana fermion since operators such as $\bar\chi\gamma^{\mu}\chi \equiv 0$. More generically one may also consider bosonic dark matter.

\item In the \MFV ansatz, we saw from the chiral structure that scalar couplings naturally follow the mass hierarchy while vector couplings tend to be universal. The latter condition is not necessary even within the \MFV framework. For example, if the leading order spin-1 flavor spurion $\delta_{ij}$  were to vanish, the next-to-leading term is $y_i^\dag y_j$ which has an even strongly hierarchical coupling to the third generation. Such a structure may be possible through models of partial compositeness \cite{Contino:2011np, Bellazzini:2012tv}.

\item We limited our analysis to a single class of mediator at a time. In the presence of multiple mediator fields, one can find processes that violate the relation between diagram topology and partial wave. For example, $\chi\bar\chi \to \varphi_1 \varphi_2$ is $s$-wave for distinct spin-0 particles $\varphi_{1,2}$.
\end{itemize}

\section{Conclusions and Outlook}

\begin{table}
\begin{center}
\begin{tabular}{cccccccclcl}
    \toprule 
    &
    \multicolumn{2}{l}{\textsc{Mass} [\textsc{g}e\textsc{v}] } &
    \, &
    \multicolumn{2}{l}{\textsc{Interaction}} &
    \quad &
    \multicolumn{2}{l}{\textsc{Coupling}} &
    &
    \textbf{Thermal}
    \\
    \textbf{Mediator} & 
    $m_\chi$ & 
    $m_\text{mes.}$ & 
    &
    \textbf{{\small DM}} & 
    \textbf{{\small SM}} & 
    &
    $\lambda_\text\DM$ &
    $\lambda_\text\SM$ &
    &
    \textbf{Relic?}
    \\
    \hline
    \rule{0pt}{2ex}
    spin-0 &
    110 & 
    20 &
    &
    $\gamma^5$ &
    $\mathbbm{1}$ & 
    &
    1.2 &
    $< 0.08$ 
    &
    &
    \textsc{msp}?
    \\
    \textquotedbl &
    \textquotedbl & 
    \textquotedbl &
    &
    $\gamma^5$ &
    $\gamma^5$ &
    &
    \textquotedbl &
    $<0.02^*$ 
    &
    &
    \textquotedbl
    \\
    spin-1 &
    45 & 
    14 &
    &
    $\gamma^\mu$ &
    $\gamma_\mu$ &
    &
    0.18 & 
    $< 10^{-6}$
    &
    &
    $\gamma=1.3$
    \\
    \textquotedbl &
    \textquotedbl & 
    \textquotedbl &
    &
    $\gamma^\mu\gamma^5$ &
    $\gamma_\mu\gamma^5$ &
    &
    \textquotedbl &
    $< 0.004$
    &&\textquotedbl
    \\
    \textquotedbl &
    \textquotedbl & 
    \textquotedbl &
    &
    $\gamma^\mu\gamma^5$ &
    $\gamma_\mu$ &
    &
    \textquotedbl &
    $< 0.006$
    &&\textquotedbl
    \\
    \textquotedbl &
    \textquotedbl & 
    \textquotedbl &
    &
    $\gamma^\mu$ &
    $\gamma_\mu\gamma^5$ &
    &
    \textquotedbl &
    $< 0.02$
    &&\textquotedbl
    \\
    \bottomrule 
\end{tabular}
\caption{Best fit parameters assuming $b$-philic couplings for the spin-0 mediator and universal quark couplings for the spin-1 mediator. The upper bound for $\lambda_\text\SM$ for the $\gamma^5\otimes\gamma^5$ is a conservative estimate for the 8 \TeV mono-$b$ reach at the \LHC (see Section~\ref{sec:collider}); the other bounds come from direct detection.
In the last column, we indicate whether consistency with a thermal relic abundance suggests a tighter \DM profile ($\gamma=1.3$) or some population of millisecond pulsars (\textsc{msp}), see Section~\ref{sec:relic:abundance}.
}\label{table:conclusion}
\end{center}
\end{table}

We have presented a class of simplified models where dark matter annihilates into on-shell mediators which, in turn, decay into the \SM with a typically suppressed width. This separates the sector of the model which can account for indirect detection signals---such as the \FERMI galactic center $\gamma$-ray excess---and those which are bounded by direct detection and collider experiments. We have addressed $\gamma$-ray spectrum coming from these models and have compared used the $\gamma$-ray excess to identify plausible regions of parameter space for a \DM interpretation; the best fit parameters and bounds on the \SM coupling are shown in Table~\ref{table:conclusion}. We have addressed the key points for \UV model building and, in an appendix below, highlight further directions for modifying the $\gamma$-ray spectrum with more general \SM final states.

\section*{Acknowledgements}

This work is supported in part by the \textsc{nsf} grant \textsc{phy}-1316792. 
We thank 
Kev Abazajian, 
Nikhil Anand, 
Nicolas Canac, 
Eugenio Del Nobile, 
Jonathan Feng, 
Shunsaku Horiuchi, 
Manoj Kaplinghat, 
Gordan Krnjaic (`\emph{kern-ya-yitch}'), 
Tongyan Lin, 
Simona Murgia, 
Brian Shuve, 
Tracy Slatyer, 
Yuhsin Tsai, 
Daniel Whiteson, 
and 
Hai-Bo Yu 
for many insightful discussions. 
Plots in this document were generated using \textit{Mathematica} \cite{Mathematica8}. \textsc{p.t.}\ would like to thank Southwest airlines and the airspace above the California coast where part of this work was completed. 

\vspace{1em}

\noindent While this paper was being prepared, \cite{Boehm:2014bia, Ko:2014gha} was posted with an explicit model for on-shell vector mediators. \cite{Boehm:2014bia} differs from the $\chi\bar\chi \to VV$ mode in this work in that it examines a specific \UV completion which includes semi-annihilations. Their $1\sigma$ contours also do not account for the systematic uncertainties discussed in Sec.~\ref{sec:fitting:spectra}. Shortly after this work was posted to ar\textsc{x}iv, \cite{Martin:2014sxa} was posted and explores on-shell mediators with diverse \SM final states and emphasizes the theme in our Figs.~\ref{fig:ID:4q}--\ref{fig:SM:pollution} and Appendix~\ref{app:spectral:shape} that one need not focus only on bottom quark couplings and, further, that dark matter masses both above and below 40 \GeV can yield the $\gamma$-ray excess.

\appendix

\section{The Spectrum of Spectra}
\label{app:spectral:shape}

In the main text we have shown how the conventional 40 \GeV \DM model for the $\gamma$-ray excess can be converted into a heavier \DM model ($m_\chi = 80$, $120$ \GeV) by taking the limit where annihilation to on-shell mediators dominates. We further showed that one can interpolate the \DM masses between $m_\chi= 40 \GeV$ and 80, 120 \GeV by parametrically increasing the \SM coupling and increasing the fraction annihilations through an off-shell mediator.
In this appendix we briefly demonstrate nonstandard (i.e.\ beyond $b\bar b$ and $\tau\bar\tau$) spectra that may also fit the $\gamma$-ray excess in the regimes $m_\chi < 40$ \GeV and $m_\chi >$ 80, 120 \GeV.
We use \PPPC as described in Sec.~\ref{sec:generate:gam:rays:methodology} and our fits are subject to  the caveats described in Sec.~\ref{sec:fitting:spectra}. For simplicity and consistency when comparing to other plots in this paper, we plot the data fit to the $b\bar b$ template from Fig.~8 of \cite{Abazajian:2014fta}.

\begin{figure}

    \begin{subfigure}[b]{0.47\textwidth}
        \includegraphics[width=\textwidth]{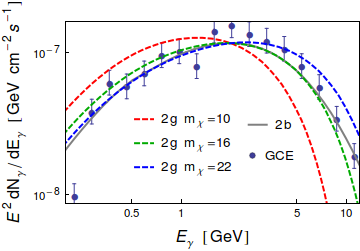}
        \caption{}
    \end{subfigure}
    \begin{subfigure}[b]{0.47\textwidth}
        \includegraphics[width=\textwidth]{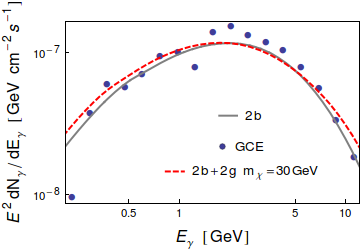}
        \caption{}
    \end{subfigure}

    \begin{subfigure}[b]{0.47\textwidth}
        \includegraphics[width=\textwidth]{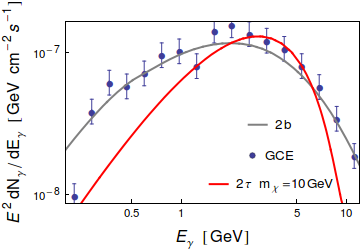}
        \caption{}
    \end{subfigure}
    \begin{subfigure}[b]{0.47\textwidth}
        \includegraphics[width=\textwidth]{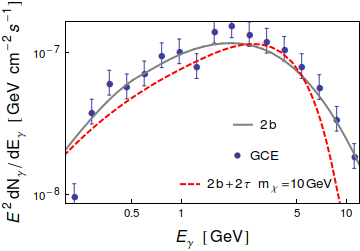}
        \caption{}
    \end{subfigure}

    \begin{subfigure}[b]{0.47\textwidth}
        \includegraphics[width=\textwidth]{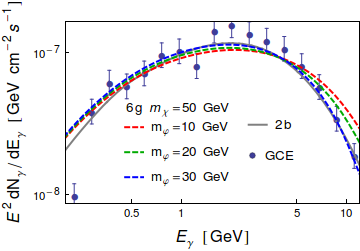}
        \caption{}
    \end{subfigure}
    \begin{subfigure}[b]{0.47\textwidth}
        \includegraphics[width=\textwidth]{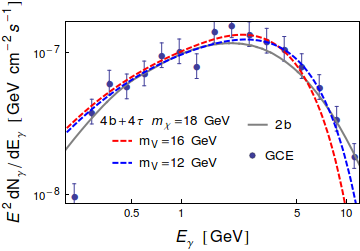}
        \caption{}
    \end{subfigure}

\caption{(a) $\chi\bar\chi \to gg$, (b) $\chi\bar\chi \to gg \text{ (67\%) or } b \bar b\text{ (33\%)}$, (c) $\chi\bar \chi \to \tau\bar\tau$, (d) $\chi\bar \chi \to \tau\bar\tau \text{ (85\%) or } b \bar b\text{ (15\%)}$, (e) $\chi\bar\chi \to 6g$, (f) $\chi\bar\chi \to 2\times\left[\tau\bar\tau \text{ (85\%) or } b\bar b\text{ (15\%)}\right]$. Spectra for various final states, including branching ratios to different final states. 4-(6-)body final states originate from on-shell mediators with masses $m_V$ ($m_\varphi$) shown. For visual comparison with other plots in this work, the gray $2b$ line is the $\chi\bar\chi\to b\bar b$ best fit spectrum and dots are the measured galactic center $\gamma$-ray excess spectrum (\textsc{gce}) assuming a $b\bar b$ signal template from \cite{Abazajian:2014fta}. Bars demonstrate an arbitrary measure of goodness-of-fit with respect to this spectrum. Note that the $\gamma$-ray excess data depends on the template used for the \DM $\gamma$-ray spectrum so these data points are mainly for comparative purposes and are not necessarily representative of the goodness-of-fit to the $\gamma$-ray excess. See Sec.~\ref{sec:fitting:spectra} for details. }
\label{fig:crazy:spectra}
\end{figure}


Fig.~\ref{fig:crazy:spectra} shows sample spectra that show the range of behavior when considering different final states both for off-shell $s$-channel processes and for those with on-shell mediators. In each of these cases, we note that by considering either admixtures of different final states or on-shell mediator annihilation into different species, one can find viable \DM models for the $\gamma$-ray excess where the \DM mass is less than the 40~\GeV value typically considered in the literature.

For example, we point out in (a) and (b) that gluons can give a reasonable fit to the spectrum. While the photon spectrum from monochromatic gluons takes a slightly different shape than that of the $b$---presumably part of the reason why $gg$ final states were not proposed for the $\gamma$-ray excess fit---they are reasonably close to the data given the implicit systematic uncertainties. This fit is improved significantly if the [off-shell, $s$-channel] mediator is allowed to decay to both gluons or $b\bar b$ pairs. Shown in (b) is the fit for a mediator that decays to either gluons or $b\bar b$ pairs, with
\begin{align}
\text{Br}(\text{mediator}\to gg) \approx 2\,  \text{Br}(\text{mediator}\to b\bar b).
\end{align}
The gluon mode is especially amenable to lighter dark matter masses since the final state is massless. Couplings to a spin-0 mediator can be generated through, for example, loops of third generation quarks.

Similarly, in Fig.~\ref{fig:crazy:spectra}(c) we show what appears to be a poor fit to 10 \GeV $\tau\bar\tau$ pairs. This, however, is a consequence of comparing the $\gamma$-ray spectrum from $\tau\bar\tau$ to the $\gamma$-ray excess fit assuming a $b\bar b$ \DM template. It is indeed well known that \DM annihilating into 10 \GeV $\tau$s fits the excess well; this should be taken as a reminder of the systematic uncertainties implicit with the $\gamma$-ray fits. It also serves to highlight that for a specific model, a proper assessment of the fit to the $\gamma$ ray excess requires a full astrophysical fit to the specific annihilation mode (along the lines of \cite{Abazajian:2010zy} and \cite{Daylan:2014rsa}) where both the model parameters and background parameters are fit simultaneously. For our purposes here, we only highlight the change in the spectrum from (c) to (d) where we introduce a 15\% branching ratio of the mediator going to $b\bar b$---the fit has interpolated between the two spectra and gives an intuitive handle for how to generate hybrid spectra. A similar hybrid spectrum was explored in Fig.~6 of \cite{Macias:2013vya}.

In Fig.~\ref{fig:crazy:spectra}(e, f) we demonstrate the range of behavior for annihilation to on-shell mediators that each decay to either gluons or $\tau\bar\tau$/$b\bar b$.  Note that an on-shell vector mediator cannot decay into two gluons by the Landau-Yang theorem so that one is forced to consider either $\chi\bar\chi \to 2\times (V\to ggg)$ or $\chi\bar\chi \to 3\times(\phi \to gg)$, each with six final state gluons. We plot the latter case in (e). In (f) we see an example of an on-shell vector mediator that decays to $\tau\bar\tau$ 85\% of the time and $b\bar b$ the remainder. This spectrum fits the $\gamma$-ray excess spectrum for a $b\bar b$ template with $m_V \approx 12$ \GeV.

Finally, we propose a simple extension where the \DM mass can be made heavier than the region considered in the primary text. We saw that the on-shell mediator scenario raised the \DM mass by having \DM annihilation go into more final state primaries ($b$ quarks). By extending the mediator sector to include additional on-shell states between the \DM and \SM sectors in Fig.~\ref{fig:simplified:model}, one may force larger dark matter masses. For example, \cite{Mardon:2009rc} explored the cascade where $\chi\bar\chi \to 2\phi_1$ with $\phi_i \to 2 \phi_{i+1}$ for the \textsc{Pamela} positron excess \cite{Adriani:2008zr}. See the appendix in that paper for analytical results for the generalization of the box spectrum to a higher polynomial spectrum where the degree of the polynomial is set by the number of on-shell mediator sectors. Additionally, as we mentioned above, one may use the Landau-Yang theorem to force $V_1 \to 3g$ decays at the end of the cascade or use mediator sectors where symmetries force $\phi_i \to n \phi_{i+1}$ with $n>2$. We remember from our analysis in Sec.~\ref{sec:relic:abundance}, however, that increasing the number of on-shell mediators per annihilation while maintaining the $\gamma$-ray excess signal also increases the annihilation cross section beyond what is expected from a simple thermal relic.

\scriptsize

\bibliographystyle{utphys} 
\bibliography{HeavyHooperon}

\end{document}